\documentclass[12ppt,twocolumn]{emulateapj}
\usepackage{amsmath,amssymb,rotfloat}
\usepackage{graphics}
\usepackage{graphicx}
\usepackage{color}
\usepackage{ulem}
\usepackage[section]{placeins}

\newcommand{\pder}[2]{\frac{\partial #1}{\partial #2}}

\newcommand{\jsc}{J. Sci. Comp.}
\newcommand{\jcap}{J. Cosm. Astropart. Phys. }
\begin{document}

\shorttitle{Magnetic field amplification in collisionless plasmas}
\shortauthors{Falceta-Gon\c calves \& Kowal}
\title{Fast magnetic field amplification in the early Universe: growth of collisionless plasma instabilities in turbulent media}

\author{D. Falceta-Gon\c{c}alves\altaffilmark{1,2} \& G. Kowal\altaffilmark{2}}
\altaffiltext{1}{SUPA, School of Physics \& Astronomy, University of St Andrews, North Haugh, St Andrews, Fife KY16 9SS, UK}
\altaffiltext{2}{Escola de Artes, Ci\^encias e Humanidades, Universidade de S\~ao Paulo, Rua Arlindo Bettio, 1000, S\~ao Paulo, SP 03828-000, Brazil}

\begin{abstract}
In this work we report a numerical study of the cosmic magnetic field
amplification due to collisionless plasma instabilities. The collisionless magnetohydrodynamic equations
 derived account for the pressure anisotropy that leads, in specific
conditions, to the firehose and mirror instabilities. We study the time
evolution of seed fields in turbulence under the influence
of such instabilities. An approximate analytical time evolution of magnetic field is provided. 
The numerical simulations and the analytical
predictions are compared. We found that i) amplification of magnetic field was
efficient in firehose unstable turbulent regimes, but not in the mirror unstable
models, ii) the growth rate of the magnetic energy density
is much faster than the turbulent dynamo, iii) the efficient amplification
occurs at small scales. The analytical prediction for the correlation between
the growth timescales with pressure anisotropy ratio is confirmed by the
numerical simulations. These results reinforce the idea that pressure
anisotropies - driven naturally in a turbulent collisionless medium, e.g. the
intergalactic medium -, could efficiently amplify the magnetic field in the early
Universe (post-recombination era), previous to the collapse of the first
large-scale gravitational structures. This mechanism, though fast for the small scale fields ($\sim$kpc scales), is however unable to provide relatively strong magnetic fields at large scales. Other mechanisms that were not accounted here (e.g., collisional turbulence once instabilities are quenched, velocity shear, or gravitationally induced inflows of gas into galaxies and clusters) could operate afterwards to build up large scale coherent field structures in the long time evolution.
\end{abstract}


\section{Introduction}

The baryonic fraction of the intergalactic medium (IGM) is composed by dillute warm/hot plasmas. Depending on the 
local environment the plasma properties of the IGM may be different. For instance, the most diffuse component of the IGM, known as intergalactic ``voids", are vast regions between clusters of galaxies in the Universe as seen today. 
Filaments of somewhat compressed and warmer ($10^5-10^7$K) gas form what is broadly understood as the IGM, while the denser ($\sim 10^{-3}$cm$^{-3}$) and shock heated ($\sim 10^7-10^8$K) plasma in clusters of galaxies are known as the intracluster medium (ICM). For many decades the intracluster medium is known to be magnetized
\citep[see reviews by][and references
therein]{kronberg94,ensslin2005,widrow12,durrer13}, but unfortunately any contraints on the 
magnetization of the IGM and intergalactic voids lack of more substantial 
observational confirmation. 
The magnetic intensity and its
respective correlation length must naturally be related to the
processes from which they originate and/or amplify, which depend on
 local dynamical and physical properties.
However, the origin and/or amplification of the magnetic fields at
cosmological scales are not completely understood yet, and represent a long
standing major issue in modern astrophysics.

Seed fields may have been generated at the pre-equipartition epoch by subatomic scale processes, such
as hadron phase transitions, or at later (post-equipartition) and pre-galactic epoch by large-scale
magnetohydrodynamical (MHD) processes, e.g. the ${\bf \nabla}p \times {\bf
\nabla}\rho$ battery term from Ohm's Law \citep{biermann50}, 
currents induced by electronic scattering of anisotropic radiation fields \citep{langer05,durrive14}, 
evolution of turbulent fields \citep[e.g.][and many others]{banerjee04}, and even shock excited Weibel 
instabilities \citep{schlickeiser03,schlickeiser05}. \citep{jeda98} suggests however that photon diffusion damping would affect post-equipartition, but pre-recombination, magnetic fluctuations. Therefore most of the amplification processes mentioned above would only be important after recombination era. Theese 
could in principle generate magnetic fields with intensities
of the order of $10^{-30}-10^{-11}$G. There are several implications of a strong magnetic field on the
cosmological evolution of the Universe, from primordial nucleosynthesis to the
statistics of the cosmic microwave background (CMB). Current data limits the
primordial (comoving) magnetic field intensity to $B_{\rm 1Mpc} < 5$nG at 1Mpc
lengthscale \citep[see][for references on WMAP and PLANCK data analysis,
respectively]{trivedi10,planck14}. Such field intensities are below the values inferred for the ICM ($B\sim 0.1-10~\mu$G \citep[see][and references therein]{govoni_feretti_2004}), as well as for more diffuse groups of galaxies \citep[e.g.][]{nikiel13}. Recently, arguable lower limits for the magnetization of the diffuse intergalactic medium have been also provided based on gamma-ray observations of blazars \citep[see][]{nero10}. These evidences combined together suggest that the Universe was magnetized even before the structure formation. 
Therefore, the problem seems to be twofold: the
first being the actual origin of the magnetic fields (seed fields), which may have occured very early
in the Universe history (pre-equipartition); and the second being the amplification of such seed
fields to the values observed in the diffuse gas of clusters of galaxies.

Observations of the Faraday rotation (FR) effect provide currently the best
estimates for the IGM magnetic fields. In the FR effect the position angle of
linearly polarized radiation shifts on the plane of sky as a function of
wavelength $\lambda$ as $\Delta \phi = {RM} \times \lambda^2$, with the rotation
measure $RM$ being $RM({\rm rad/m^{2}}) \simeq 812 \int_0^{L({\rm kpc})}n_{\rm
e}({\rm cm^{-3}}) B_\parallel(\mu{\rm G}) dl$, where $\parallel$ denotes the
direction parallel to the line of sight. 

It is well-known that RM strongly depends on the spatial distribution of the
electron density ($n_e$) and the magnetic field geometry along the line of sight
($B_\parallel$), and basic geometrical assumptions with respect to them are
usually made. It is also natural to presume that the IGM magnetic field is not
uniform. Faraday rotation maps present ordered magnetic fluctuations at scales
from $\sim 1$kpc associated with cooling flows up to $\sim 1$Mpc associated to
powerful active galactic nuclei (AGN) jets and cluster radio relics
\citep[e.g.][]{ferreti2012}. These lengthscales, however, cannot be directly
understood as correlation scales for the magnetic field. Observational estimates
of typical lengthscales usually assume an {\it ad hoc} correlation between the
magnetic field and local plasma density in the range of $1-30$kpc
\citep[e.g.][]{bonafede10,vacca12}. Unfortunately, given the turbulent nature of
the IGM, the actual correlation length ($l_{\rm corr}$) of the IGM/ICM magnetic
field depends also on the spatial distribution of the velocity field
\citep[e.g.][]{burk09,xu2009,falceta14}, and  $l_{\rm corr}$ for the magnetic
field of the IGM cannot be observationally estimated yet. Still, even though around equipartition 
level with the thermal and kinetic counterparts, the presence of strong magnetic
fields ($B \gtrsim \mu$G) is surprising, given the general 
understanding that $\mu$G scale fields would be the result of galactic 
dynamo amplification of much less intense pre-galactic seed fields and 
the related timescales.

During the last decade several authors have employed numerical simulations and analytical 
approximate solutions of collisional plasmas to study
the magnetic amplification due to turbulent dynamo, galactic fields diffused
into the ICM by outflows, and AGNs \citep[see e.g.][and references
therein]{dolag02,dubois09,donnert09,falceta10,xu2012,schober12, schober13,cho14,federrath14}. It was found that all
these processes could in principle account for the amplification of a
pre-existing seed field to observable values. Such degeneracy among different
mechanisms could be removed if the correlation lengths of density, velocity and
magnetic fields were directly and independently obtained observationally. This because 
the different processes suggested so far operate at either different scales, or 
with different cross-correlations between the physical parameters (${\bf B}$,$n$,$p$,${\bf v}$).
However, the possibility that $\mu$G magnetic fields could be present in vast volumes of
the more diffuse intergalactic medium, where the density of galaxies and the
impact of AGNs are relatively small, points to the turbulent dynamo as a viable
and ubiquitous process.

Great effort on the understanding of turbulent dynamos has been employed by
means of analytical and numerical studies \citep[see][and references
therein]{schek07,brandenburg12,schober12,bete13}. The turbulent dynamo is the process by
which kinetic energy of turbulent motions is converted into magnetic energy.
Turbulent cells stretch and fold field lines in a weakly magnetized plasma
increasing the total magnetic energy. Such a dynamo can in principle operate at
the large range of scales where there is turbulence, depending on the properties of the flow. This was first pointed
out by \cite{batch50}, and further developed into the Kraichnan-Kazantsev
theory \citep{kaz67,krai68}. Here, and throughout this paper, we assume the intergalactic
plasma to be a highly conducting medium in the sense that the resistivity
($\eta$) is much smaller than the viscosity ($\nu$), i.e. the Prandtl number $Pm
\equiv \nu/\eta \gg 1$. In this case, the small-scale dynamo (SSD) theory predicts an exponential growth of
magnetic energy, up to equipartition at the viscous scale. The power spectrum of
the magnetic fluctuations is then proportional to $k^{3/2}$, resulting in a
concentration of energy at small scales. 
Once saturated at small scales, the dynamo enters in a nonlinear phase showing
slower amplification rates (linear with time). The scale at which turbulence is dissipated  
regulates the efficiency of nonlinear phase for the dynamo process. If we consider the 
standard collisionality of the ICM, the non-linear phase of the
turbulent dynamo is possibly of little interest for cosmological magnetic field studies. 
It has been suggested though that the enhanced 
effective collisionality due to plasma instabilities could even completely suppress 
the kinetic phase, and the nonlinear dynamo would be all that is left 
\citep[see][]{mogavero14}.  
As we discuss below, for the purpose of this work, the exact determination of the 
transition between kinetic\footnote{when the
magnetic energy density is small compared to the kinetic energy density of the
smallest turbulent scale} and nonlinear regimes is not relevant. Due to the little 
understanding of the anomalous collisionality in diffuse media, and the fact that the 
nonlinear regime is even slower than the kinetic one, we will 
concentrate only on the early kinetic phase.

We may consider that during the kinetic phase the magnetic field intensity
increases as $B(t) = B_0 \exp\left(t/\tau_{\rm d}\right)$, being the timescale
$\tau_{\rm d} = l_{\rm d}/\delta v_{l_{\rm d}}$, with $\delta v_{l_{\rm d}}$ the
turbulent amplitude at the viscous scale $l_{\rm d}$. The saturation occurs when
$B(t_{\rm sat}) \simeq \delta v_{l_{\rm d}} \left(4\pi \rho\right)^{1/2} \simeq
\delta v_{\rm L} Re^{-1/4} \left(4\pi \rho\right)^{1/2}$, which results
in\footnote{Here we define the Reynolds number as $Re = \delta v_L L/\nu$, being
$\nu$ the fluid viscosity and $L$ the largest turbulent scale. We also make use
of the Kolmogorov scaling relation for the velocity $\delta v_{\rm l} \propto
l^{1/3}$}:

\begin{equation}
t_{\rm sat} \sim 2  \frac{L}{\delta v_{\rm L}} Re^{-1/2} \ln\left({\cal M}_{\rm A,L}^0 Re^{-1/4} \right),
\label{eq:sattime}
\end{equation}
\noindent
where $\rho$ stands for the mass density of the fluid, and ${\cal M}_{\rm A,L}^0
= \delta v_{\rm L} \left(4\pi \rho\right)^{1/2} / B_0$ the Alfv\'en Mach number
at the large scale $L$ with respect to the seed field intensity $B_0$. Large
turbulent Reynolds numbers result in faster SSDs, however the saturation occurs
at lower amplification levels reducing its effectiveness. In the opposite trend,
relatively smaller values of $Re$ result in slower SSDs, but with larger
$B(t_{\rm sat})$. However, in order to explain the cosmological amplification of
magnetic fields after recombination era we need large $B(t_{\rm sat})$ and fast
amplification. During the formation of clusters of galaxies, starting at at $z
\sim 1.0$ (which represents a {\it look back time} of $< 8$Gyrs\footnote{For a
standard $\Lambda$CDM cosmological model, assuming a dark energy density $\Lambda = 0.714$,
matter density (baryons + dark matter) $\Omega_{\rm M} = 0.286$ and a Hubble parameter
$h=0.7$}),
velocities of the order of hundreds of km/s\footnote{as estimated from
equipartition with thermal component, i.e. mildly transonic turbulence} are
driven at scales as large as $L \sim 1$Mpc. For typical intergalactic gas
densities of  $n \sim 10^{-3}$cm$^{-3}$ one finds $B(t_{\rm sat}) \sim \mu$G for
$Re \sim 10 - 1000$, which resembles the values expected for the IGM using the
Spitzer viscosity. For a seed field $B_0 \lesssim $nG, using the same
parameters, one also finds $t_{\rm sat} > 10$Gyrs. It is clear, at least from
these crude estimates, that only superestimated values of the magnetic seed
field and a fine tunned Reynolds number could explain the magnetic field
intensities observed in the local Universe \citep[see e.g.][]{cho14}.

Given the inefficient amplification of the field by the turbulent dynamo,
another mechanism must be considered. It has been recently pointed that the IGM
may not be well described by a standard MHD theory. The intergalactic plasma,
extremely rarefied, behaves as a gyrotropic collisionless plasma, i.e. the
Larmor radius of the ions ($r_{\rm L} \sim eB/mcv_{\rm th,i}$) is smaller than the mean free path
$\lambda_{\rm mfp} \sim (n \sigma_{\rm ii})^{1/2}$, being $\sigma_{\rm ii}$ the ion-ion collision 
cross-section \citep[][]{sche05,lima14}. This relation can be rewritten as the condition for 
gyrotropic plasmas as $B_{\rm gyro}({\rm G}) \gtrsim 10^{-20} n({\rm cm^{-3}})T({\rm eV})^{-1/2}$. 
In the standard $\Lambda$CDM Universe, with an average density of baryons at $z \sim 1000$ of $n_{z1000} \sim
10^{6}$cm$^{-3}$ and $kT_{z1000} \sim 1$eV, one finds $r_{\rm L} < \lambda_{\rm
mfp}$ for magnetic fields larger than $\sim 10^{-14}$G, which decreases with time as 
the Universe expands. Therefore, for the given seed fields above, it is natural to 
assume the intergalactic plasma as gyrotropic. If not during its whole evolution 
post-recombination evolution, at least from the time when the condition $r_{\rm L}<\lambda_{\rm mfp}$ 
is satisfied.

Even in the fluid
approximation \citep[][CGL-MHD hereafter]{chew56} gyrotropic plasmas are subject
to a number of instabilities that provide fast interchange of internal, kinetic
and magnetic energies. \citet{sche06a} presented a simplified analytical study
of the time evolution of the magnetic energy in the ICM based on such
instabilities. Their conclusion was that collisionless plasma instabilities may
result in an ``explosive'' growth of magnetic energies in very short timescales
compared to the cosmological evolution of the system. However, such idea has not
yet been extended into a full theory nor tested numerically, which is the main
goal of this work.

In the present paper we study numerically the amplification of magnetic fields
in turbulent collisionless plasmas, under the CGL-MHD formalism. In order to
achieve this goal we provide a number of direct numerical simulations of plasma
flows, with forced turbulence, threaded initially by extremely weak seed fields.
Collisionless plasma instabilities are included. The basic theoretical aspects
of such problem is discussed in Section 2. The numerical setup and algorithms
used are provided in Section 3. In Section 4 we present the main results,
followed by the Discussions and the Conclusions.

\section{Magnetohydrodynamic description of collionless plasmas}

The proper description of collisionless plasmas relies on the full calculation
of the particle dynamics, including both eletromagnetic and collisional forces.
Unfortunately such approach is of little practical use, with no analytical model
available yet. One of the alternatives for this problem relies on numerically
integrating the equation of motion of an ensemble of charged test particles
(ions), coupled to a fluid (electrons). Such approach is also called
particle-in-cell (PIC) numerical simulations. PIC simulations provide the
dynamical evolution of momentum distributions of particles as they interact with
magnetic fields and due to collisions, which allow us to directly study e.g. the
effects of collisionless plasma instabilities on the isotropization of
pressures, or on the rise of pressure anisotropies. However, the spatial
coverage of the computational domain in PIC simulations is limited to a finite
number of Larmor radii. For this reason it is not possible to study both the
evolution of the distributions of particle momenta and the system dynamics at
large scales, such in the case of the IGM, simultaneously. At first
approximation, we may consider the distribution of momenta of particles to be
Maxwellian (or bi-Maxwellian in the case of magnetized collisionless plasmas),
for which a fluid description of a pressure-anisotropic plasma is available
\citep{chew56}.

\subsection{Single fluid approximation of a plasma with pressure anisotropy}

The derivation of the CGL-MHD equations from the Vlasov-Maxwell equations is
provided, for example, in \citet{kulsrud_1983}. By neglecting heat conduction
and other possible heating/cooling sources, the one fluid CGL-MHD set of
equations for a plasma with pressure anistropy can be rewritten, in conservative
form, as:
\begin{equation*}
  \frac{\partial }{\partial t}
  \begin{bmatrix}
    \rho \\[6pt]
    \rho \mathbf{u} \\[6pt]
    \mathbf{B} \\[6pt]
    e
  \end{bmatrix}
  + \nabla \cdot
  \begin{bmatrix}
    \rho \mathbf{u} \\[6pt]
    \rho \mathbf{uu} + \Pi_{P} + \Pi_{B} \\[6pt]
    \mathbf{Bu - uB} \\[6pt]
    e \mathbf{u} + \mathbf{u} \cdot \left( \Pi_{P} + \Pi_{B} \right)
  \end{bmatrix}
  =
\end{equation*}
\begin{equation}
  =
  \begin{bmatrix}
    0 \\[6pt]
    \mathbf{f} \\[6pt]
    0 \\[6pt]
    \mathbf{f \cdot v}
  \end{bmatrix}
  \rm{,}
\label{eqn:collisionless_mhd}
\end{equation}
where $\rho$, $\mathbf{u}$, $\mathbf{B}$, $p_{\perp,\parallel}$ represent the
plasma density, velocity, magnetic field, and perpendicular/parallel pressures
with respect to orientation of the local magnetic field, respectively, and $e =
(p_{\perp} + p_{\parallel}/2 + \rho u^{2}/2 + B^{2}/2)$ is total energy density.
The anisotropy in pressure for a gyrotropic plasma is then defined as $\Delta
\equiv (p_{\perp} - p_{\parallel})/p_{\perp}$. $\Pi_P$ and $\Pi_B$ are the
pressure and the magnetic stress tensors, respectively, given as:
\begin{equation}
\Pi_{P} = p_{\perp} \mathbf{I} + (p_{\parallel} - p_{\perp}) \mathbf{bb} \rm{,}
\end{equation}
and
\begin{equation}
\Pi_{B} = (B^{2}/8 \pi) \mathbf{I} - \mathbf{BB} /4 \pi \rm{,}
\end{equation}
where $\mathbf{I}$ is the unitary dyadic tensor and $\mathbf{b} = \mathbf{B} /
B$. The source term $\mathbf{f}$, in the equations above, represents the
external force responsible for driving the turbulence.

The set of equations given in Eq.\ref{eqn:collisionless_mhd} is not closed.
Since the perpendicular and parallel pressures can evolve differently from each
other, another equation relating the time evolution of $p_\perp$ and
$p_\parallel$ is needed. The original double-adiabatic condition proposed in
\citet{chew56}, based on the conservation of first and second magnetic
invariants, leads to:

\begin{equation}
\frac{d }{d t} \left( \frac{p_{\perp}}{\rho B} \right) = 0, \;\;\;\;\;
\frac{d }{d t} \left( \frac{p_{\parallel} B^{2}}{\rho^{3}} \right) = 0.
\label{eqn:cgl_closure}
\end{equation}

The CGL-MHD double-adiabatic closure equations above may not be exactly correct.
It is known that several processes, e.g. collisions or enhanced particle
scattering by magnetic mirrors (or firehoses \citep[e.g.][]{kunz14}), break these invariants.
Another possible closure for Eq.\ref{eqn:collisionless_mhd} is obtained by
simply assuming that the system reaches a pressure anisotropy equilibrium fast
enough, so $\Delta = const.$ during the whole evolution. The closure problem is
discussed in more details in the following sections.

\subsection{Wave modes and stability condition in gyrotropic plasmas}

A linear perturbation analysis of Eq.\ref{eqn:collisionless_mhd} results in the
known modified dispersion relations in gyrotropic plasmas given below \citep[see
e.g.][]{hau_wang_2007,kowal11}:

\begin{equation}
\left( \frac{\omega}{k} \right)^{2}_{a} =
\left( \frac{B^{2}}{4 \pi \rho} +
\frac{p_{\perp}}{\rho} -
\frac{p_{\parallel}}{\rho} \right)
\cos^{2}\theta,
\label{eqn:cgl_alfven_speed}
\end{equation}
corresponding to the Alfv\'en transversal mode, with $\cos \theta = \mathbf{k
\cdot B}/(kB)$ (being $\mathbf{k}$ the wavevector of the perturbation), and:

\begin{equation}
\left( \frac{\omega}{k} \right)^{2}_{f,s} =
\frac{b \pm \sqrt{b^{2} - 4 c}}{2},
\label{eqn:cgl_fastslow_speed}
\end{equation}
corresponding the fast (``+'') and slow (``-'') magnetosonic waves, being

\begin{equation*}
b = \frac{B^{2}}{4 \pi \rho} +
\frac{2 p_{\perp}}{\rho} +
\left(\frac{2 p_{\parallel}}{\rho} -
\frac{p_{\perp}}{\rho} \right)
\cos^{2}\theta ,
\end{equation*}
\begin{equation*}
c = - \cos^{2}\theta
\left[
\left( \frac{3 p_{\parallel}}{\rho} \right)^{2} \cos^{2}\theta -
\frac{3 p_{\parallel}}{\rho} b +
\left( \frac{p_{\perp}}{\rho} \right)^{2}
\sin^{2}\theta
\right] .
\end{equation*}

The equations above result in the following stability conditions:

\begin{eqnarray}
|\Delta| \equiv \frac{|p_{\perp}-p_{\parallel}|}{p_{\perp}} > 2\beta_{\perp}^{-1}   {\rm , \;\;\;\; for  \;\;\;\;} (p_{\parallel}>p_{\perp}) , \nonumber \\[6pt] {\rm and \;\;\;\;\;\;} \nonumber \\[6pt]
\frac{1}{6}\frac{p_{\perp}}{p_{\parallel}} > 1 + \beta_{\perp}^{-1}  {\rm , \;\;\;\; for  \;\;\;\;} (p_{\parallel}<p_{\perp})
\label{eqn:firehose_crit}
\end{eqnarray}
which correspond to the firehose and fluid mirror instabilities, respectively. We must point out that 
there is a factor of $1/6$ for the later, which is not present in the threshold of the actual 
mirror instability obtained from the kinetic theory. This offset is a well-known issue in the CGL-MHD closure, but this is not relevant for this work as we focus on the evolution of the firehose mode only.

\section{An approximate analytical solution for the amplification of $B$ in the unstable regime}

By differentiating the pressure anisotropy $\Delta$ in time one obtains:

\begin{equation}
\label{anisotropy}
\frac{d \Delta}{dt} = \frac{1}{p_\perp} \left[ \left(1-\Delta \right)  \frac{d p_\perp}{dt} - \frac{d p_\parallel}{dt}\right].
\end{equation}

The first term of the right side of Eq.\ref{anisotropy} can be evaluated in
terms of the magnetic field intensity. The magnetic moment, or first adiabatic
invariant, $\mu = m_i v_\perp^2/2B$ {\rm = constant}, of a charged particle
subject to the magnetic force, is also quasi-invariant for a plasma. Therefore,
by approximating $v$ to the thermal velocity $v_{\rm th}$,
associated to the Maxwellian distribution of velocity components perpendicular to
the magnetic field, and assuming $d \rho/\rho \ll dB/B$, we find $dp_\perp/p_\perp \simeq dB/B$, and
Eq.\ref{anisotropy} results in:

\begin{equation}
\label{anisotropy2}
\frac{d \Delta}{dt} \simeq \left( 1-\Delta \right) \frac{1}{B}\frac{dB}{dt} + \mathcal{F}_{\rm iso},
\end{equation}

\noindent
where

\begin{equation}
\mathcal{F}_{\rm iso} = -\frac{1}{p_\perp}\frac{dp_\parallel}{dt}-\nu_{\rm ii} |\Delta|,
\end{equation}

\noindent
represents an effective isotropization rate ($\nu_{\rm eff} \sim
\mathcal{F}_{\rm iso} \Delta^{-1}$). Notice that the first term in the right
hand side of the equation above corresponds to the time evolution of $p_\parallel$ 
as a consequence of the changes in $B$ and $p_\perp$. From here we describe 
the interchange of parallel and perpendicular pressures, mediated by magnetic fluctuations,
 as the net magnetic scattering of the particle distributions ($\nu_{\rm scatt} \sim p_\perp^{-1} \Delta^{-1} dp_\parallel/dt$). The last term of the equation above has been deliberately added to account for the effect of
collisions, by means of a Braginskii collision frequency $\nu_{\rm ii}$. From
these two terms it is possible to consider two different regimes with: i)
collision dominated isotropization, for $\nu_{\rm ii} \gg \nu_{\rm scatt}$; and
ii) scattering dominated isotropization, for $\nu_{\rm ii} \ll \nu_{\rm scatt}$.

The effect of collisionless instabilities on the pitch angle (magnetic)
scattering is not fully understood. If we take the firehose instability as
example, perturbations (e.g. in the velocity field) may result in local
growth of magnetic field intensity ($\delta B$). This in turn results in an increase of
the perpendicular velocities of the ions, i.e. an increase of $p_\perp$,
at the expense of a decrease of $p_\parallel$. This could in principle
be used to estimate $dp_\parallel/dt$\footnote{e.g. as with the use of the 2nd adiabatic 
invariant}. However the total internal energy may not
be conserved if particles are lost during scattering, or if other energy loss 
processes are included, making the problem not
practical without a full kinetic description. An alternative, as pointed by
\citet{sche06b}, is to assume this term to saturate  the amplification of the
magnetic field perturbations quasi-linearly at the wavelength of fastest growth,
which occurs around the Larmor radius ($\omega \sim \Omega_{\rm i}$, being
$\Omega_{rm i} = eB/m_{\rm i}c$ the cyclotron frequency of the ions) \citep{gary98},
therefore $\delta B/B|_{\rm sat} \sim \gamma_{\rm
max} \Omega_{\rm i}^{-1}$, given
the maximum growth rate (see Eqs.\ref{eqn:cgl_alfven_speed} and
\ref{eqn:cgl_fastslow_speed}):

\begin{equation}
\gamma_{\rm max} \sim \left( \left|\Delta \right| - \frac{2}{\beta} \right)^{\alpha}\Omega_{\rm i},
\end{equation}

\noindent
with $\alpha = 1$ and $1/2$ for mirror and firehose instabilities, respectively.
As we show further in this paper the firehose instability is of major interest
in the amplification of $B$, and we use $\alpha = 1/2$ from here. The scattering
frequency can then be estimated as $\nu_{\rm scatt} \simeq \delta B^2/B^2
\gamma_{\rm max} \sim (|\Delta|-2/\beta)^{3/2} \Omega_{\rm i}$, and
Eq.\ref{anisotropy2} becomes:

\begin{equation}
\label{anisotropy3}
\frac{d \Delta}{dt} \sim \left( 1-\Delta \right) \frac{1}{B}\frac{dB}{dt} + \left(|\Delta|-\frac{2}{\beta} \right)^{3/2}\Omega_{\rm i} - \nu_{\rm ii}|\Delta|.
\end{equation}

\subsection{Magnetic field amplification with constant $\Delta$}

This problem is tractable analytically if we assume that the evolution of the
firehose unstable regions in the plasma occur with constant $\Delta$. This
approximation is justified by the fact that any process that may amplify the
pressure anisotropy (e.g. turbulence, anisotropic cosmic ray scattering, and
others) is precisely counterbalanced by the ones leading to isotropization. This
is actually a probable scenario given that the increase of the pressure
anisotropy by external sources is independent on $\Delta$, while the
isotropization processes are a function of the anisotropy itself. Therefore, the
system should evolve around an ``equilibrium'' anisotropy value $\Delta_0$.
If the relaxation timescale is short compared to the dynamical timescales of the
system, we may assume $\Delta(t) \simeq \Delta_0$. We must stress that the validity 
of such conjecture depends on the level of anisotropy itself, 
i.e. it is probably correct for $\Delta_0 \rightarrow 0$ and certainly does not stand 
 for $\Delta_0 \gg 0$.

\subsubsection{the case of $\nu_{\rm ii} \gg \nu_{\rm scatt}$}

Let us first consider the case in which collisions dominate the isotropization
of pressure. This limit could well describe the initial stages of the evolution
of a cosmological seed field after the recombination era, given that $\beta \gg
1$ and $\nu_{\rm ii} \gg \Omega_{\rm i}$. The evolution of the magnetic field in
this case depends mostly on the  dynamics of the plasma, and not on
instabilities driven by the pressure anisotropy. Given that the resistive
dissipation is extremely small in the intergalactic medium we may apply the
``frozen-in'' condition to the plasma. The evolution of the magnetic field is
then dominated by the turbulent rate of strain at the viscous scales \citep[see
e.g.][]{sche06b}:

\begin{equation}
\label{dyn1}
\frac{1}{B} \frac{dB}{dt} \simeq {\bf b} {\bf b} \cdot{\bf \nabla u} - {\bf \nabla \cdot u} \sim \frac{\delta v_{\rm L}}{L} Re^{1/2}.
\end{equation}

\noindent
For $\Delta = {\rm const}$ and $\nu_{\rm ii} \gg \nu_{\rm scatt}$, and combined
with Eq.\ref{dyn1}, Eq.\ref{anisotropy3} is then reduced to:

\begin{equation}
\label{turbanis}
|\Delta| \sim \left(\frac{v_{\rm th}}{\delta v_{\rm L}}\right)^{3/2} \left( \frac{L}{\lambda_{\rm mfp}} \right)^{1/2} -1.
\end{equation}

\noindent
As discussed in the previous section the kinetic phase of the turbulent dynamo
(Eq.\ref{dyn1}) provides an exponential growth of the seed field up to
equipartition at the smallest turbulent scales. However, as $B$ grows,
$\Omega_{\rm i} \rightarrow \nu_{\rm ii}$ and the plasma becomes gyrotropic.
This is supposed to occur for extremely weak magnetic fields ($B_{\rm gyro}({\rm G}) \gtrsim 10^{-20} n({\rm cm^{-3}})T({\rm eV})^{-1/2}$), possibly even earlier
than the saturation of the turbulent dynamo at small scales.

Another interesting aspect of the turbulence during the early evolutionary phase
is to generate the pressure anisotropy by itself
(Eq.\ref{turbanis}). This is physically understood as the consequence of a fast
growth of the magnetic field not counterbalanced by collisions. Depending on the
properties of the turbulence, one obtains $|\Delta| \gg 0$. Such values are not
expected in highly magnetized plasmas. Once $\nu_{\rm ii} \ll \nu_{\rm scatt}$,
due to the amplification of $B$, the balance is reached since $\nu_{\rm scatt}$
strongly depends on the magnetic field intensity. Local measurements of 
the Earth's magnetosphere and the solar wind reveal pressure anisotropies around its 
quasi-stability threshold $|\Delta| \sim |\Delta|_{\rm crit}$, 
and many signatures of its role on the dynamical evolution of these systems have been 
found \citep[e.g.][]{winter95,soucek08,bale09}. The interesting implication of
Eq.\ref{turbanis} is that, due to its lower magnetization level (much higher $\beta$), 
$|\Delta|$ in the IGM/ICM could have been
much larger at higher redshits, compared to the values found presently in the
Solar System.

\subsubsection{the case of $\nu_{\rm ii} \ll \nu_{\rm scatt}$}
\label{sec_bevol}

Once the magnetic field becomes strong enough, for the plasma to become
gyrotropic, instabilities will take over the amplification of the magnetic
field. Now, as $\Delta = {\rm const}$ and $\nu_{\rm ii} \ll \nu_{\rm scatt}$,
and replacing the cyclotron frequency $\Omega_i$, Eq.\ref{anisotropy3}
becomes\footnote{notice that the amplification of $B$ in our model can occur for
firehose instability, i.e. $p_\parallel > p_\perp$ (see Eq.\ref{anisotropy}),
and therefore $\Delta<0$ and $1-\Delta > 1$}:

\begin{equation}
\label{bevol}
\frac{dB}{dt} \sim \frac{e}{m_{\rm i}c}  \frac{\left( |\Delta|-2\beta^{-1}\right)^{3/2}}{1+|\Delta|} B^2,
\end{equation}

\noindent
which, at $\beta \gg |\Delta|^{-1}$ limit, result in a first order nonlinear
ordinary differential equation with an analytical solution:

\begin{equation}
B(t) \sim \left(B_0^{-1} - At \right)^{-1},
\end{equation}

\noindent
with $A \simeq \frac{e}{m_{\rm i}c}\frac{|\Delta|^{3/2}}{1+|\Delta|}$. The
magnetic field growth is explosive around $t \sim A^{-1}$, and the saturation
occurs for $B(\tau_{\rm gr}) \gg B_0$. The quasi-stability condition, given by
$\beta \sim 2/|\Delta|$ level, is reached in a timescale:

\begin{equation}
\label{growthrate}
\tau_{\rm gr}{\rm (s)} \sim ( |\Delta|^{\psi}\Omega_{i,0})^{-1} \sim 10^{-4}B_0^{-1}{\rm (G)} |\Delta|^{-\psi},
\end{equation}

\noindent
where $\Omega_{i,0}$ represents the cyclotron frequency measured for the seed
field $B_0$, and $\psi = 1/2$, or $3/2$, for $|\Delta| \gg 1$ and $|\Delta| \ll
1$, respectively. Eq.\ref{growthrate} then provides an approximate timescale for
the explosive growth of the magnetic field due to the collisionless plasma
instability. For example, if we consider a seed field of $B_0 \sim 10^{-17}$G
subject to such instabilities, with $|\Delta|>10^{-2}$, one finds amplification
up to equipartition with pressure anisotropy ($\beta \sim |2\Delta|^{-1}$) at $t
< 400$Myrs.

\begin{figure}
\centering
\includegraphics[width=8cm]{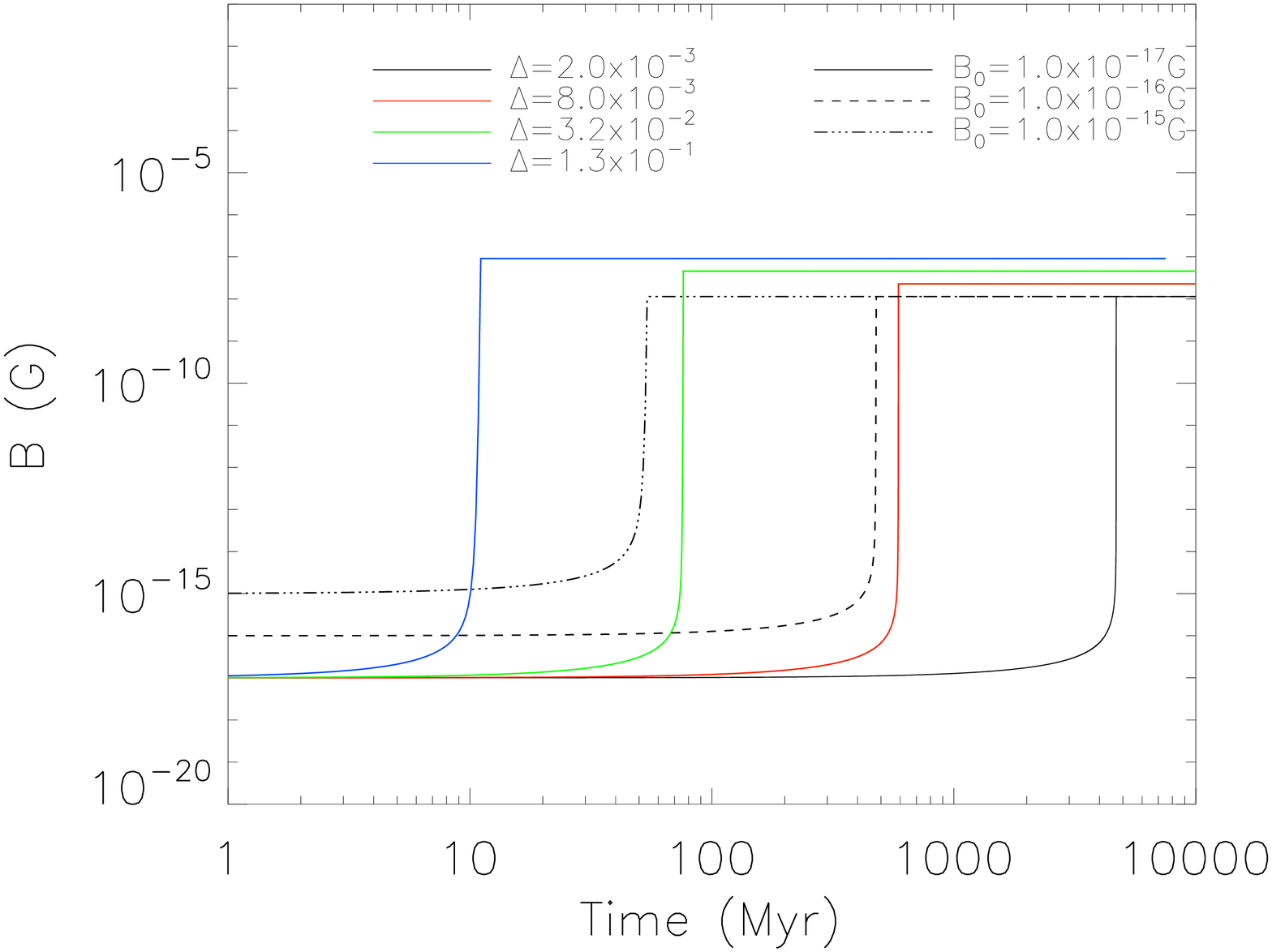}
\caption{Time evolution of magnetic field amplitude in the approximate case of
Eq.\ref{bevol}, as described in Section \ref{sec_bevol}.}
\label{fig_bevol}
\end{figure}

As pointed before, such values of pressure anisotropies are easily reached in
turbulent weakly magnetized collisionless plasmas. The main conclusions of this
section are that: i) pressure anisotropies could be naturally generated in the
IGM/ICM in the early Universe (post-recombination era) at
timescales comparable to the eddy turnover time at viscous dissipation scales,
and ii) the instabilities driven by the pressure anisotropy . A full numerical
integration of Eq.\ref{bevol} is shown in Figure \ref{fig_bevol}, which confirms
the main results obtained in the approximate limits $\beta \gg |\Delta|^{-1}$
and $B(\tau_{\rm gr}) \gg B_0$.

\section{Numerical simulations}

In the previous section we demonstrated that the evolution of plasma
instabilities driven by a pressure anisotropy could, in principle, explain the
amplification of a magnetic field seed at recombination era to the values
observed in the local Universe. These estimates, however, were obtained under
strict approximations and limit cases. In this section we describe the numerical
methods used for modelling of the dynamical evolution of the unstable plasmas.

\begin{table*}[!hbt]
\caption{Initial and saturation properties of the numerical simulations}
\centering
\begin{tabular}{c | c  c  c  c  c c | c c}
\hline \hline
 model & res & ${\cal M}_{A,0}$ \tablenote{${\cal M}_{A,0} = \delta V^{\rm inj}/V_{A,0}$ is the Alfv\'en Mach number of the turbulence injected with respect to the initial magnetic field, i.e at $t=0$} & $\beta_0$ & $p_\perp$ & $p_{\parallel}$ & $|\Delta|$\tablenote{$\Delta=(p_{\perp}-p_{\parallel})/p_{\perp}$} & $\left< \beta^{\rm sat} \right>$ & $\tau_{\rm sat}$($t_{\rm dyn}$) \\
[0.5ex]
\hline
fh512Ma0D1.0 & $512^3$ & 2.0 & 2.0 & 0.25 & 1.0 & 3.0 & 1.6  & 0.52 \\
mr512Ma0D0.5 & $512^3$ & 1.0 & 3.0 & 4.0 & 1.0 & 0.75 & 4.3  & - \\
mr256Ma4D0.8 & $256^3$ & $10^4$ & $6.8 \times 10^{7}$ & 1.0 & 0.04 & 0.96 & $2.5 \times 10^7$  & -\\
mr256Ma4D0.6 & $256^3$ & $10^4$ & $7.5 \times 10^{7}$ & 1.0 & 0.25 & 0.75 & $2.2 \times 10^7$  & -\\
fh512Ma4D1.0 & $512^3$ & $10^4$ & $2.0 \times 10^{8}$ & 1.0 & 4.0 & 3.0 & 7.6  & 0.36 \\
fh256Ma4D1.0 & $256^3$ & $10^4$ & $2.0 \times 10^{8}$ & 1.0 & 4.0 & 3.0 & 7.8 &  0.62 \\
fh128Ma4D1.0 & $128^3$ & $10^4$ & $2.0 \times 10^{8}$ & 1.0 & 4.0 & 3.0 & 7.9 &  1.40 \\
fh256Ma4D4.0 & $256^3$ & $10^4$ & $9.0 \times 10^{8}$ & 1.0 & 25.0 & 24.0 & 1.4 &  0.13 \\
fh256Ma4D9.0 & $256^3$ & $10^4$ & $3.4 \times 10^{9}$ & 1.0 & 100.0 & 99.0 & 0.6  & 0.07 \\
fh256Ma6D4.0 & $256^3$ & $10^6$ & $9.0 \times 10^{12}$ & 1.0 & 25.0 & 24.0 & 1.3  & 0.18 \\
fh256Ma8D4.0 & $256^3$ & $10^8$ & $9.0 \times 10^{16}$ & 1.0 & 25.0 & 24.0 & 1.4  & 0.22 \\
[1ex]
\hline
\end{tabular}

\label{tab:sims}
\end{table*}

\subsection{Governing equations and methods}

In order to perform the numerical modeling of the plasma evolution with the
constant pressure anisotropy and turbulence injection we used the GODUNOV
code\footnote{The MHD version of the code is publicly available from the website
\url{http://amuncode.org}} which solves the CGL-MHD equations
(Eqs.~\ref{eqn:collisionless_mhd}) in the conserved form.



For simplicity $p_\parallel = a_\parallel^2 \rho$ and $p_\perp = a_\perp^2 \rho$ are expressed by the sound speeds $a_\parallel$ and $a_\perp$, parallel and perpendicular to $\mathbf{B}$, respectively. Therefore the momentum equation can be rewritten as 

\begin{equation}
 \pder{\rho \mathbf{u}}{t} + \nabla \cdot \left[ \rho \mathbf{u} \mathbf{u} + \left( a_\perp^2 \rho + \frac{B^2}{8 \pi} \right) I - \left( 1 - \alpha \right) \frac{\mathbf{B} \mathbf{B}}{4 \pi} \right] = \mathbf{f}, \label{eq:momentum}
\end{equation}
where $\alpha = (a_\parallel^2 - a_\perp^2) / V_A^2 = \frac{1}{2} \left(\beta_\parallel - \beta_\perp \right) = \frac{1}{2} \beta_\perp \left( \xi - 1\right)$  is the pressure anisotropy degree with $\xi \equiv p_\parallel /p_\perp = a_\parallel^2 / a_\perp^2$ being the pressure ratio.

The numerical integration of the modified\footnote{standard solution of CGL-MHD equations envolves the use of the first and second adiabatic invariants as closures for the problem. Here, this is slightly modified as we avoid the use of the 2nd adiabatic invariant by fixing the pressure anisotropy as constant.} CGL-MHD equations, was done in GODUNOV using the second-order shock-capturing Godunov-scheme
\cite[see][]{kowal11}. The time integration was done using the second order
Strong Stability Preserving Runge-Kutta (SSPRK) method \citep[see][and
references therein]{gottlieb09}. Spatial reconstruction was done using the
second order the total variation diminishing (TVD) interpolation with the Van
Leer limiter \citep{vanleer74}, and numerical fluxes were calculated using the
general Harten-Lax-van Leer (HLL) Riemann solver \cite[see][e.g.]{einfeldt88}.
We incorporated the field interpolated constrained transport (CT) scheme
\cite[see][e.g.]{toth00} into the integration of the induction equation to
maintain the $\nabla \cdot \mathbf{B} = 0$ constraint numerically.

\subsection{Model of turbulence}

In our numerical modeling we drive turbulence using a method described by
\cite{alvelius99}. The forcing is implemented in spectral space and concentrated
around the injection scale related to a wave vector $k_{inj}$.  We perturb a
finite number of discrete Fourier components of velocity in a shell extending
from $k_{inj}-\Delta k_{inj}$ to $k_{inj}+\Delta k_{inj}$ with a Gaussian
profile of the half width $k_c$ and the peak amplitude $\tilde{v}_f$ at the
injection scale. The amplitude of driving is solely determined by its power
$P_{inj}$. The parameters describing our forcing do not change during the
evolution of the system.

The randomness in time makes the force neutral in the sense that it does not
directly correlate with any of the time scales of the turbulent flow, and it
also determines the power input solely by the force-force correlation.  In the
models presented in this paper we use isotropic forcing.

In particular, the total amount of power input from the forcing can be set to
balance a desired dissipation at a statistically stationary state.  In order to
contribute to the input power in the discrete equations from the force-force
correlation only, the force is determined so that the velocity-force correlation
vanishes for each Fourier mode.  The procedure of reducing the velocity-force
correlation is described in \cite{alvelius99}.

In Eq.~(\ref{eq:momentum}), the forcing is represented by a function $\mathbf{f}
= \rho \mathbf{a}$, where $\rho$ is local density and $\mathbf{a}$ is random
acceleration calculated using the method described above.

\subsection{Initial and boundary conditions}

For our calculations, similar to our earlier studies \cite[see][]{kowal11}, the
initial pressures and the uniform initial field $B_0$ are the controlling
parameters. We define the Alfv\'enic Mach number of the injected turbulence as 
${\cal M}_A = \langle \delta v / c_A
\rangle$.  The angle brackets $\langle \rangle$ represent the
volume average of the parameter.  

The distribution of $\rho=1.0$, ${\bf v}=0$ and ${\bf B}=B_0{\bf \hat{x}}$, is uniform
in the whole domain. We do not set the viscosity nor the resistivity coefficients explicitly in our
models, and dissipation is determined by the numerical scheme only. Therefore the scales at which the 
dissipation starts to be important is defined by the numerical diffusivity of the scheme.

The simulation box is perfectly periodic in all directions.

\subsection{Simulated models}

For the present work we performed a number of numerical simulation with
different plasma initial parameters, as well as different stability regimes, as
given in Table \ref{tab:sims}. We also performed simulations varying the
numerical resolutions for the purpose of convergence validation.

\section{Results}

A comparison between the spatial distributions of the density and velocity
fields in CGL-MHD and standard MHD turbulent models have been provided in
\citet{kowal11} and \citet{lima14}. In those works it has been shown that
CGL-MHD models, depending on the filling factor of unstable regimes and on the
degree of pressure anisotropy itself, may lead to evident differences in the
statistics of turbulent fields when compared to standard MHD turbulence. Those
studies were performed for relatively strong magnetic fields, i.e. for
$p_{\perp} |\Delta| \gtrsim \left<B^2\right>$, and therefore the statistics of
${\bf B}$ were weakly dependent on CGL-MHD instabilities \citep[see Fig.6
in][]{lima14}. Differently to the previous works, we focused our modelling on
the weakly magnetized turbulent regimes ($p_{\perp} |\Delta| \gg
\left<B^2\right>$).

\begin{figure*}
\centering
\includegraphics[width=0.32\textwidth]{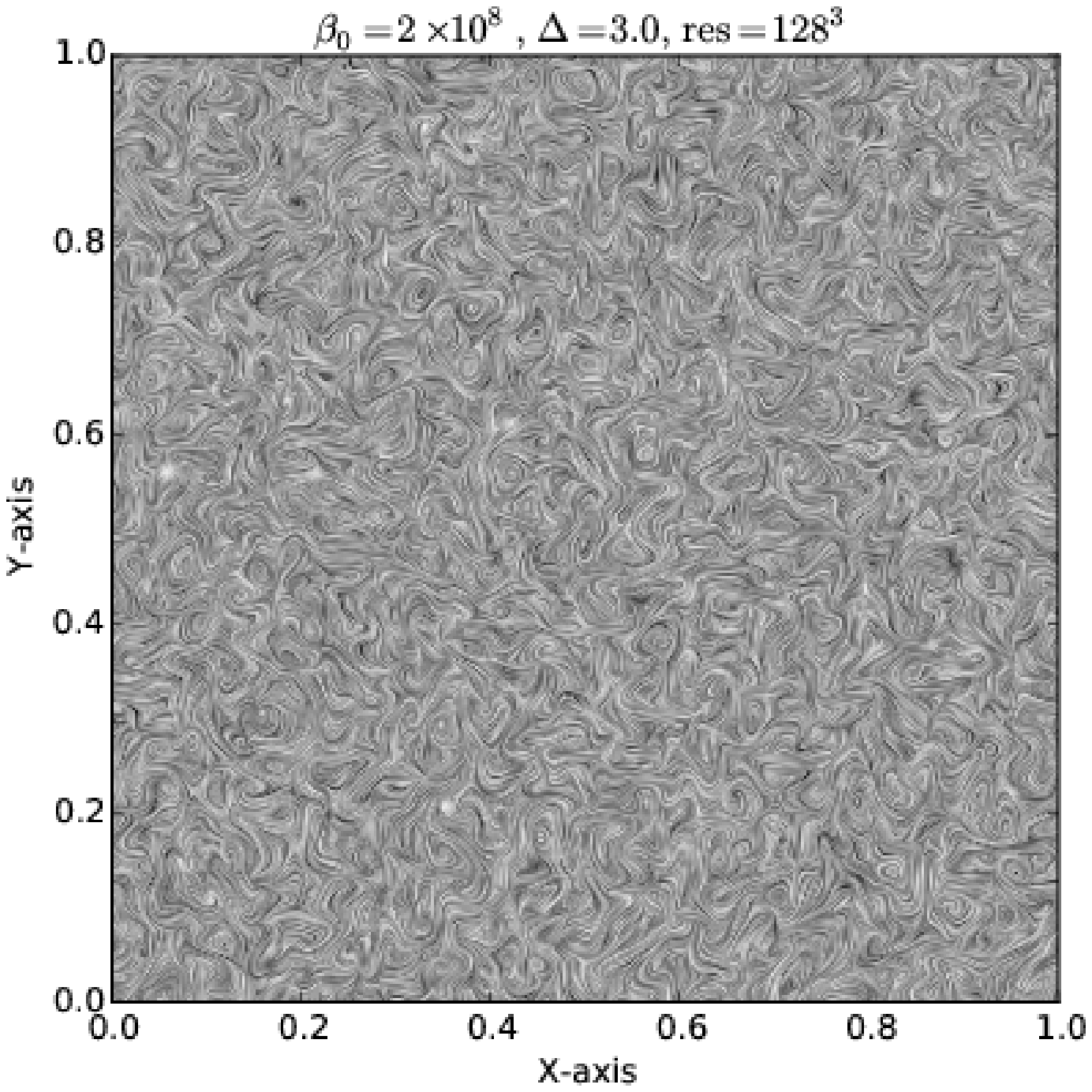}
\includegraphics[width=0.32\textwidth]{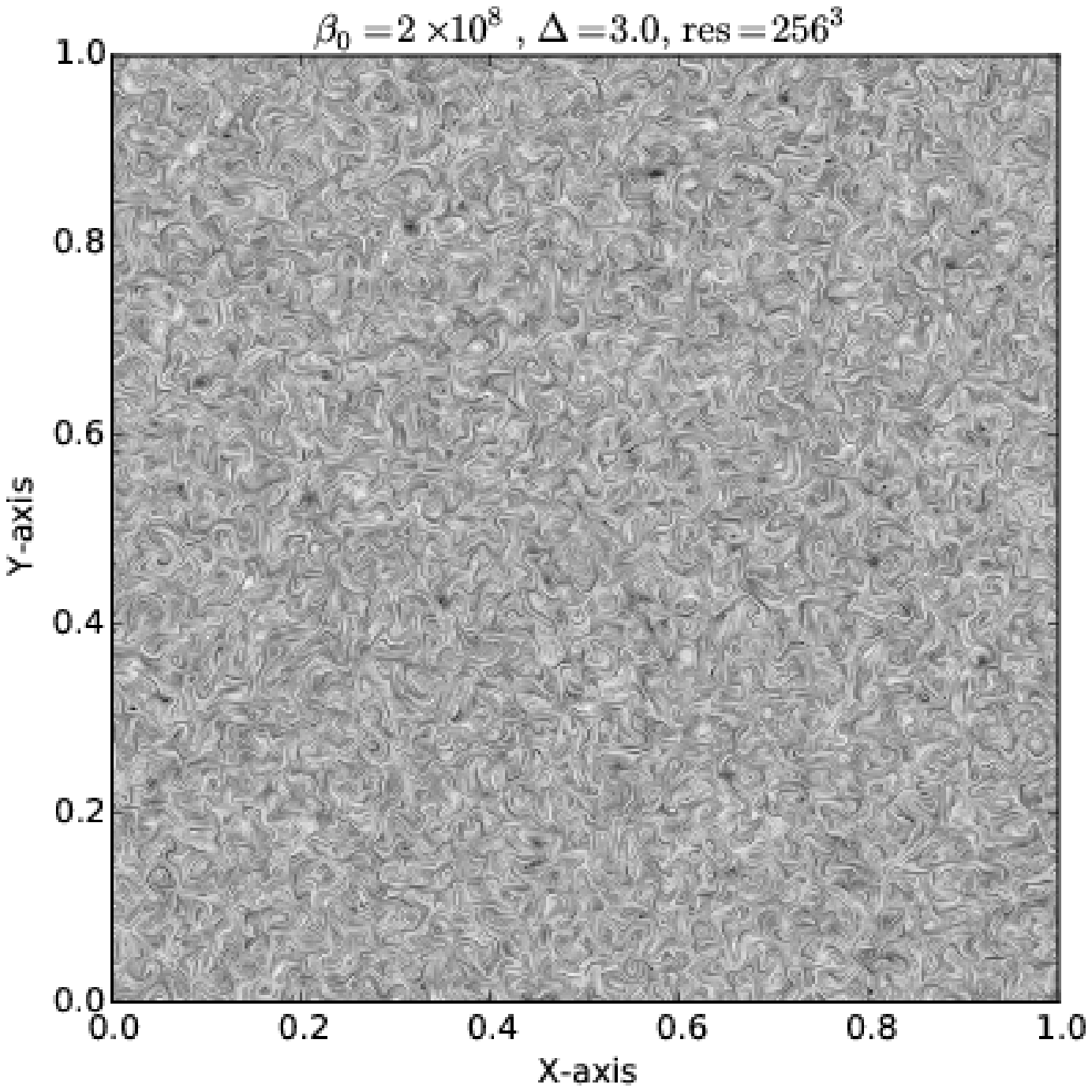}
\includegraphics[width=0.32\textwidth]{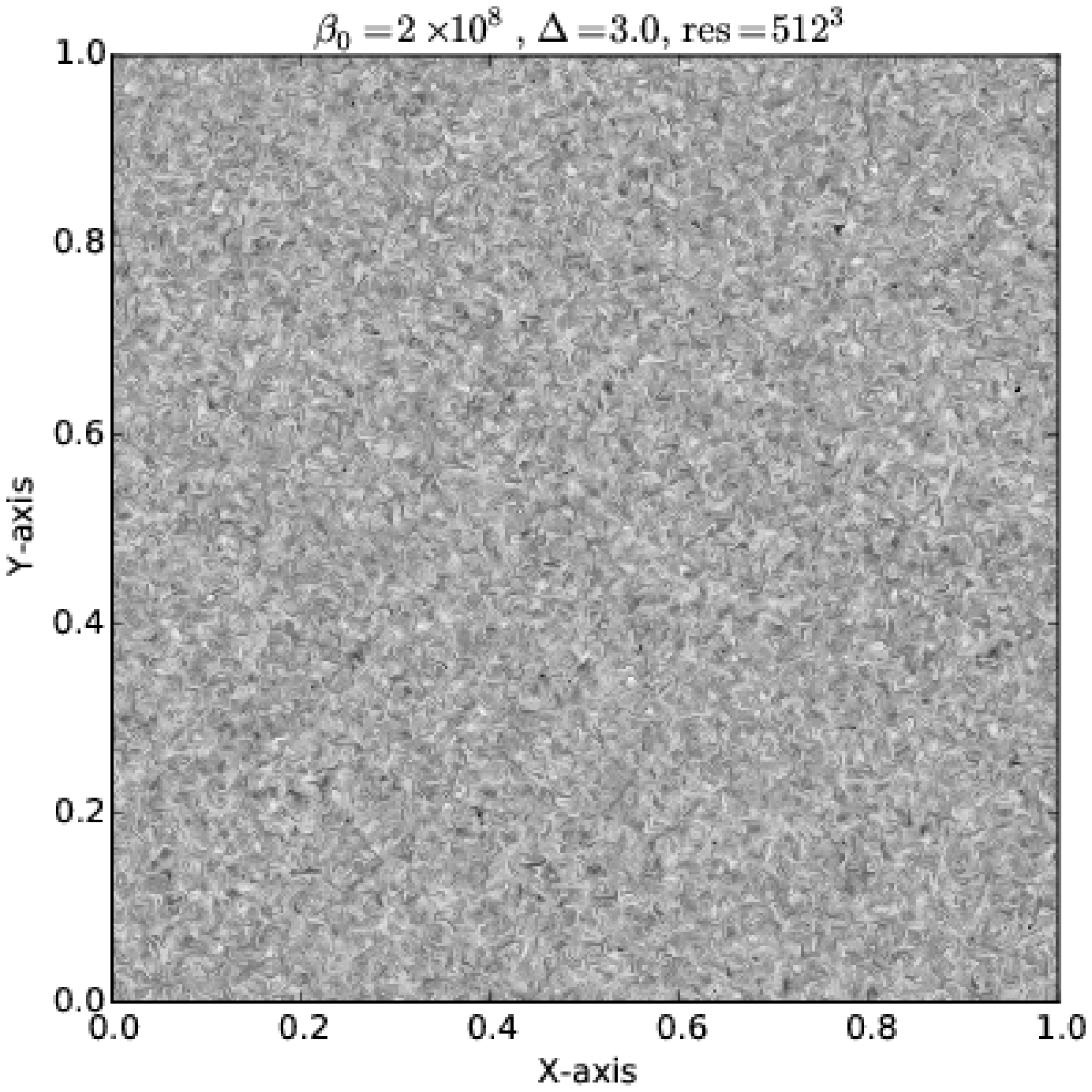}
\caption{Central slice line integral convolution (LIC) map of the magnetic
field lines at saturated stage for $|\Delta| = 3.0$ and $\beta_0 = 2 \times
10^8$, firehose unstable, models with different resolutions: $128^3$ (left),
$256^3$ (center) and $512^3$ (right).}
\label{fig_maps}
\end{figure*}

\subsection{Magnetic field struture and statistics}

As the simulations are initiated, the driving of turbulence results in
fluctuations of density, and of the velocity and magnetic fields, at different
scales. It is well known for many decades that firehose and mirror
instabilities, given the growth rate dependence with $k$
(Eqs.\ref{eqn:cgl_alfven_speed} and \ref{eqn:cgl_fastslow_speed}), result in the
emergence of structures much smaller than those observed in standard MHD
turbulence \citep{kowal11}. In standard MHD, the forcing, i.e. the process of injection of
turbulence, is the dominant dynamical process and therefore dominate the process
of structure formation. In the collisionless plasma approximation, on the other
hand, the growth rate of the instabilities rises indefinitely with the
wavenumber of the fluctuation, up to $k_{\rm max} \sim r_{\rm L}$. The numerical
scheme implemented in this work for solving the CGL-MHD equations do not account
for finite Larmor radius effects. Therefore, the maximum growth rate of the
instabilities must be related to the minimum lenghtscales of the system, which
is the size of the finite volume of the space discretization, i.e. the grid
cell. This means that both the growth rate and the geometry of the amplified
magnetic fields are resolution dependent.

In Figure \ref{fig_maps} we present the aparent configuration of the magnetic
field lines in the mid-slice (in the z-direction). The geometry of the field
lines is illustrated by the line integral convolution (LIC) technique. The LIC
imaging algorithm consists of rendering a map of random streamlines that follow
the orientation of the local field. The values are normalized to arbitrary units
in order to provide a texture map. Such texture map provides a visual
qualitative measure for the scaling of the magnetic field lines. Figure
\ref{fig_maps} shows the LIC technique applied for the the magnetic field lines
at saturated stage for $|\Delta| = 3.0$ and $\beta_0 = 2 \times 10^8$,
corresponding to firehose unstable models, for the given resolutions of $128^3$
(left), $256^3$ (center) and $512^3$ (right).

\begin{figure}
\centering
\includegraphics[width=8cm]{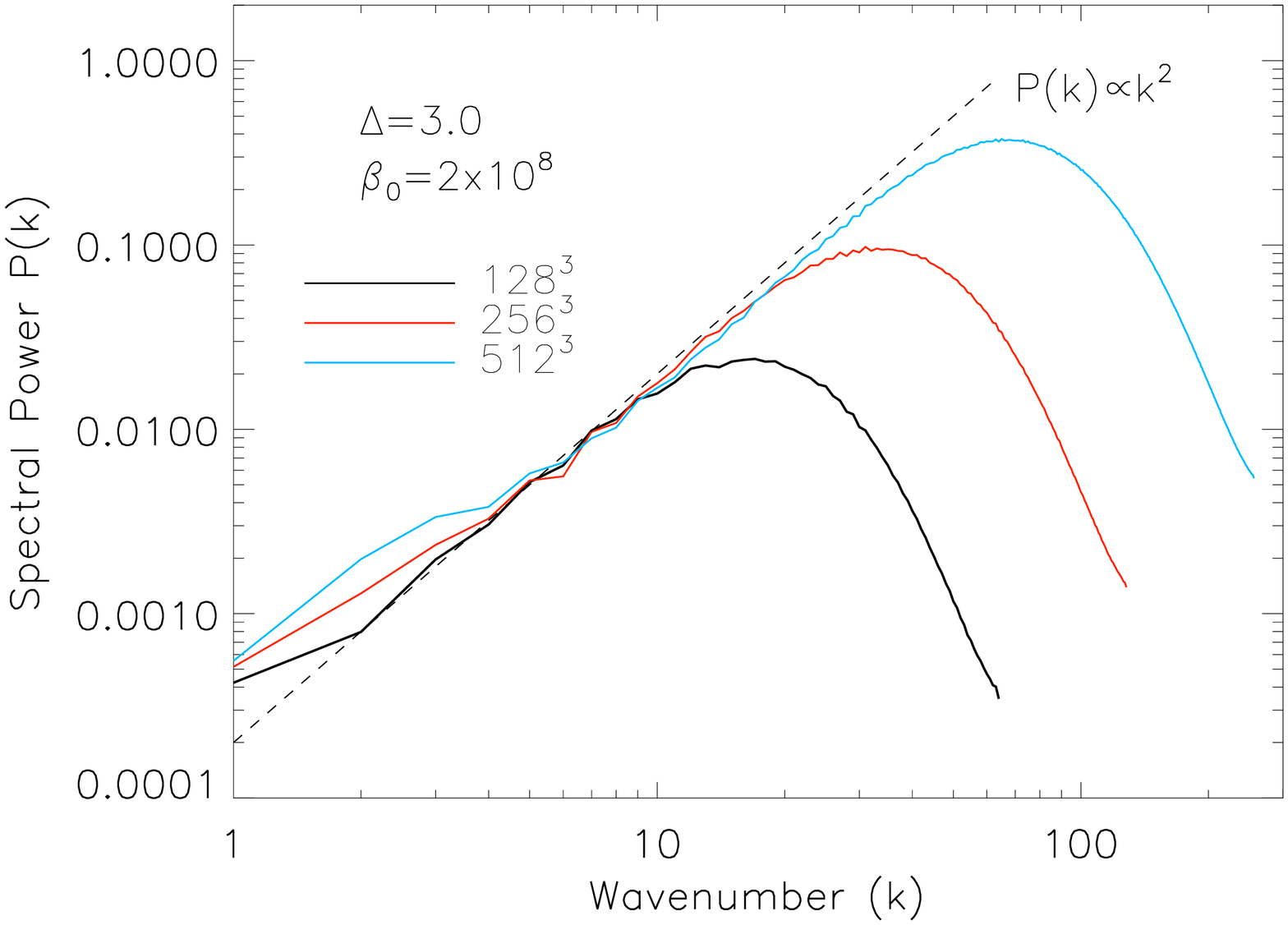}
\includegraphics[width=8cm]{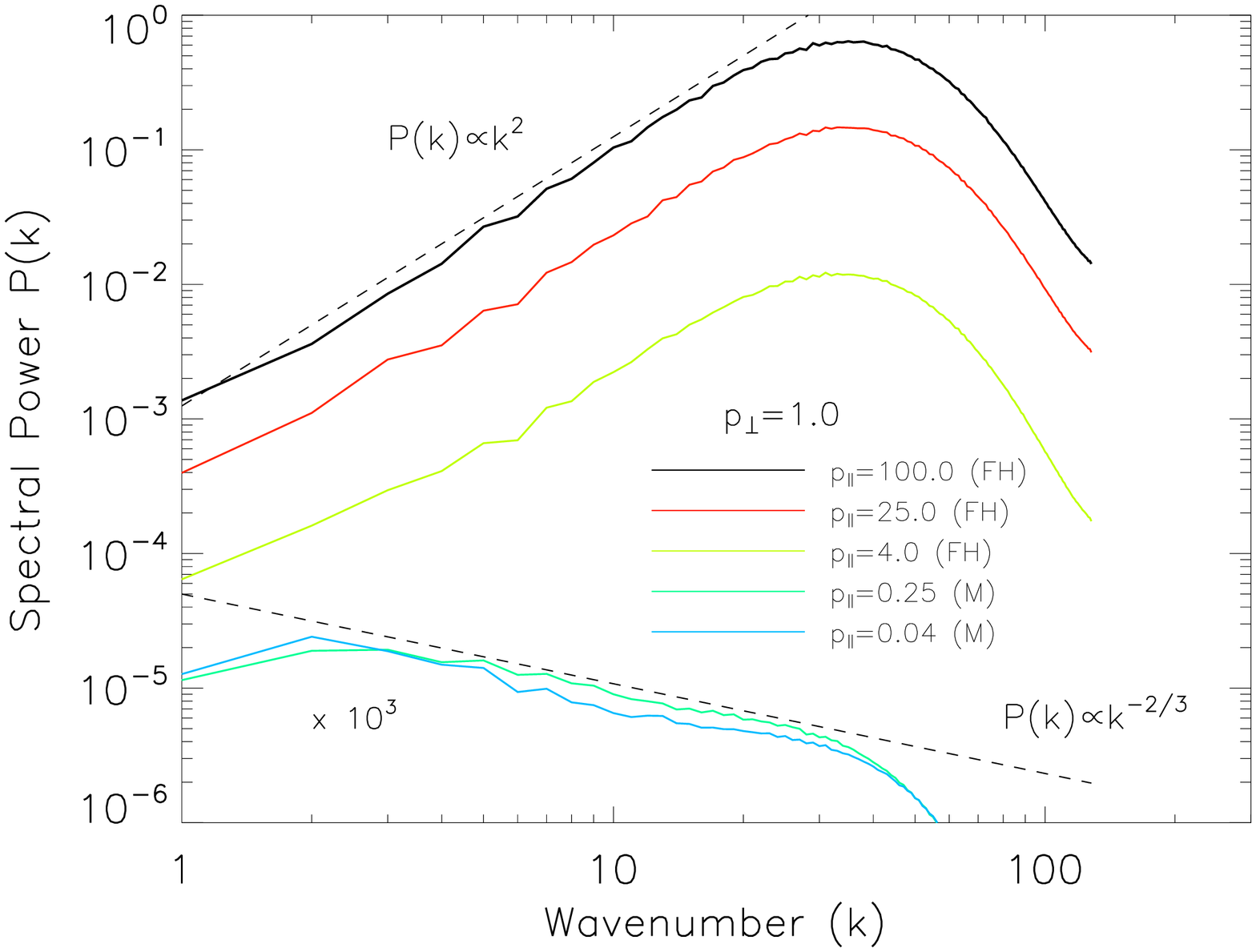}
\includegraphics[width=8cm]{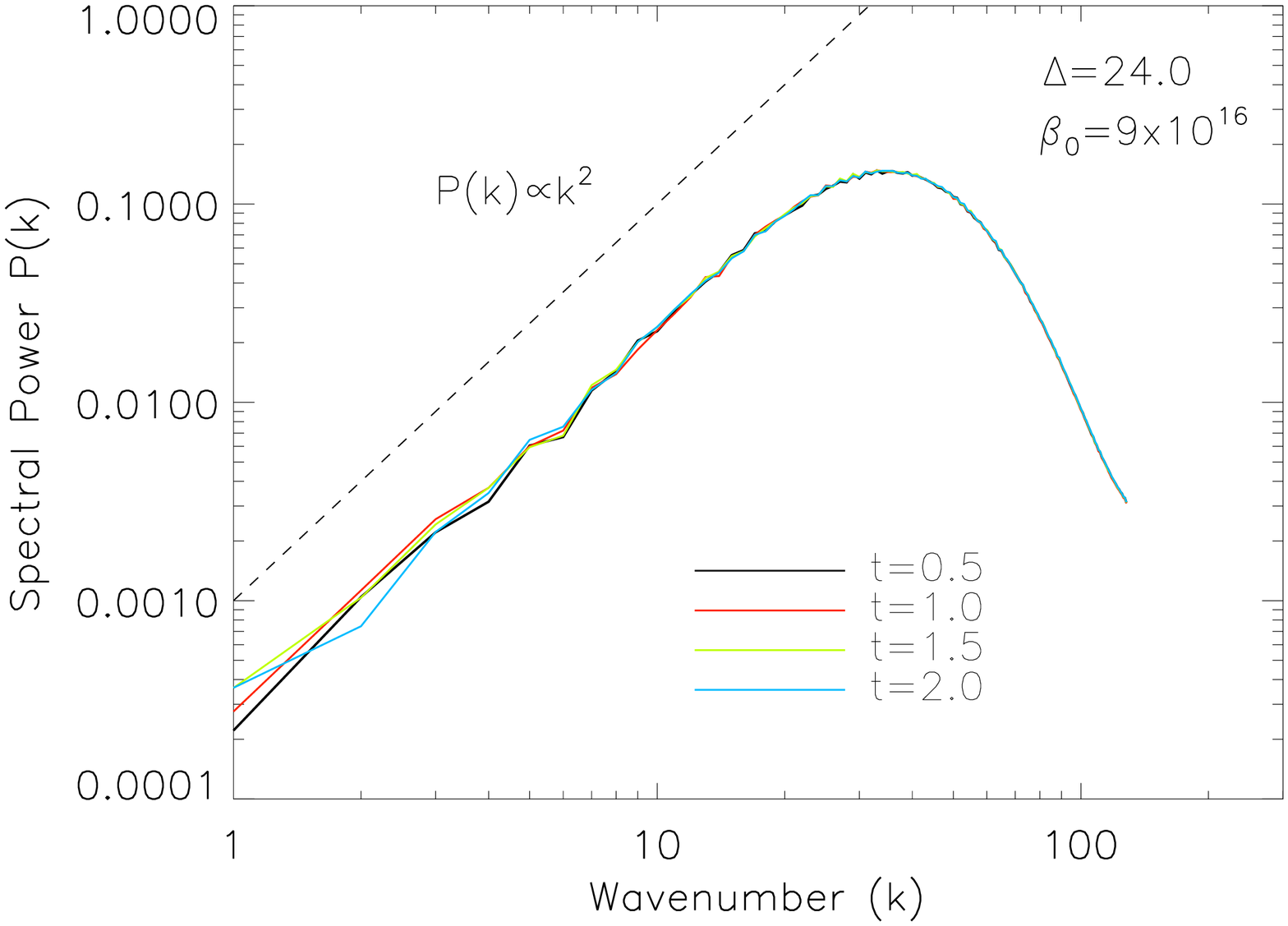}
\caption{Spatial power spectra of the magnetic field for: different numerical
resolutions (top), different pressure anisotropies (center), and at different
stages of the system evolution (bottom).}
\label{fig_spectra}
\end{figure}

It is visually clear from these plots that the finer the resolution the more
flocculent the geometry of the amplified magnetic field becomes. This result is
expected since for $|\Delta| = 3.0$ and $\beta_0 = 2 \times 10^8$ models the
thermal pressure, including the pressure anisotropy free-energy, is much larger
than the magnetic and turbulent driving energy densities. Therefore the firehose
instability dominates the plasma dynamics and the structure formation, which
growth rate should peak around the smallest scales available in the system. This
is also quantitatively measured by means of the normalized power spectrum of the
magnetic field, shown in Figure \ref{fig_spectra} (top). The three dimensional
power spectra of the magnetic field for these models reveal that the peak shifts
towards larger $k$'s as the numerical resolution increases. The peak shift is
linear with the resolution, ocurring at rounded values of $k_{\rm
peak}\simeq$16, 33 and 65, for resolutions of $128^3$ (black), $256^3$ (red) and
$512^3$ (blue), respectively. Perturbations smaller than $6-8$ grid cells suffer mostly 
the effects of numerical dissipation.

Notice that, theoretically, the maximum growth rate should occur at $k \sim
k_{\rm peak}$ but numerical diffusion is responsible for the dissipation of
coherent structures at the smallest scales. Both the numerical viscosity and
resistivity act together to reduce the effect of the instabilities in the growth
of velocity and magnetic field perturbations. If the numerical viscosity is
considered to be Laplacian we can, in principle, estimate the peak location as
the lengthscale at which the numerical viscous loss rate ($4\pi^2k_{\rm
peak}^2\nu_{\rm num}$) equals the instability grow rate, which results in
$k_{\rm peak} \sim \sqrt{|\Delta|-2\beta^{-1}} (8\pi^3 \nu_{\rm num})^{-1}$.
Given that the asymptotic values of $\beta$ depends only on $\Delta$, 
when saturation of instabilities occurs
the peak location becomes insensitive to
$(|\Delta|-2\beta^{-1})$, resulting in $k_{\rm peak} \propto \nu_{\rm
num}^{-1}$, i.e. approximately linearly proportional to the numerical
resolution as suggested before.

Also, it is noticeable that the numerical resolution, except for the location of
the peak, is not relevant for the general shape of the power spectrum of the
magnetic field fluctuations. A power spectrum $P \propto k^\alpha$, with $\alpha
\sim 2$, is observed in the range of scales $\sim [k_{\rm inj}, k_{\rm peak}]$,
in agreement with the saturation estimates for $\delta B_k^2 \propto \gamma_k^2
\propto k^2$ (see Section 3), for all the numerical resolutions. This is a good indicative of
convergence for the models. The obtained power spectrum is steeper than 
the one expected for the kinetic-phase of the small-scale dynamo, which shows 
that the instabilities are the dominant process in the field amplification here. This 
also show that the magnetic field at large scale is comparatively weaker.

The effects of $|\Delta|$ on the structure of the magnetic field is also shown
in Figure \ref{fig_spectra} (center). Here we present the spatial spectral power
of the magnetic field for the 5 different models with same thermal perpendicular
pressure ($p_\perp$) and numerical resolution ($256^3$), at $t=2.0 t_{\rm dyn}$,
but for different parallel pressures ($p_\parallel$), resulting in mirror
(``M'') and firehose (``FH'') unstable systems with different anisotropies. The
mirror unstable models, in contrast to the firehose unstable ones, present a
decreasing power with wavenumber. This is because, as shown above, the
amplification of the magnetic fluctuations in the firehose regime is dominated
by the instabilities (positive slope of the power spectra), while in the mirror unstable regions it
is basically subject to the dynamics of the flow, i.e. the turbulence
(responsible for the negative slope of the power spectra). Also, as we discuss in more details in
the next subsection, the amplitude of the power spectrum is a function of the
pressure anisotropy for the firehose unstable models, while it is insensitive to
$|\Delta|$ in the mirror unstable models. In the firehose case, the larger the
anisotropy the larger is the total power of the magnetic fluctuations, which is
related to the real values of the fluctuations and to the degree of
amplification itself.

Since saturation is an important condition for the identification of the peak
and total power of the magnetic fluctuations we studied the time evolution of
the spatial power spectrum of ${\bf B}$ of one of the firehose unstable models
($|\Delta|=24$,$\beta_0 = 9 \times 10^{16}$ and $256^3$). We chose the model
with the smallest magnetic seed field given its longer saturation timescale,
which could eventually influence the spectral power distribution. As can be seen
from Figure \ref{fig_spectra} (bottom), the spectra are very similar for all
snapshots ($t=0.5, 1.0, 1.5$ and $2.0t_{\rm dyn}$), with exactly the same slopes
and peak locations. This means that the saturation occurs at $t \ll t_{\rm
dyn}$, and that the turbulence driven - even though super-Alfv\'enic (with
respect to the seed field) - is not able to modify the spatial distribution of
the amplified field fluctuations. It is worth mentioning here that, since the 
non-linear regime (as for the small-scale dynamo) apparently does not exist, 
another mechanism is therefore required to continue the magnetic field amplification 
on larger scales.

\subsection{Magnetic field amplification}

\begin{figure*}
\centering
\includegraphics[width=8cm]{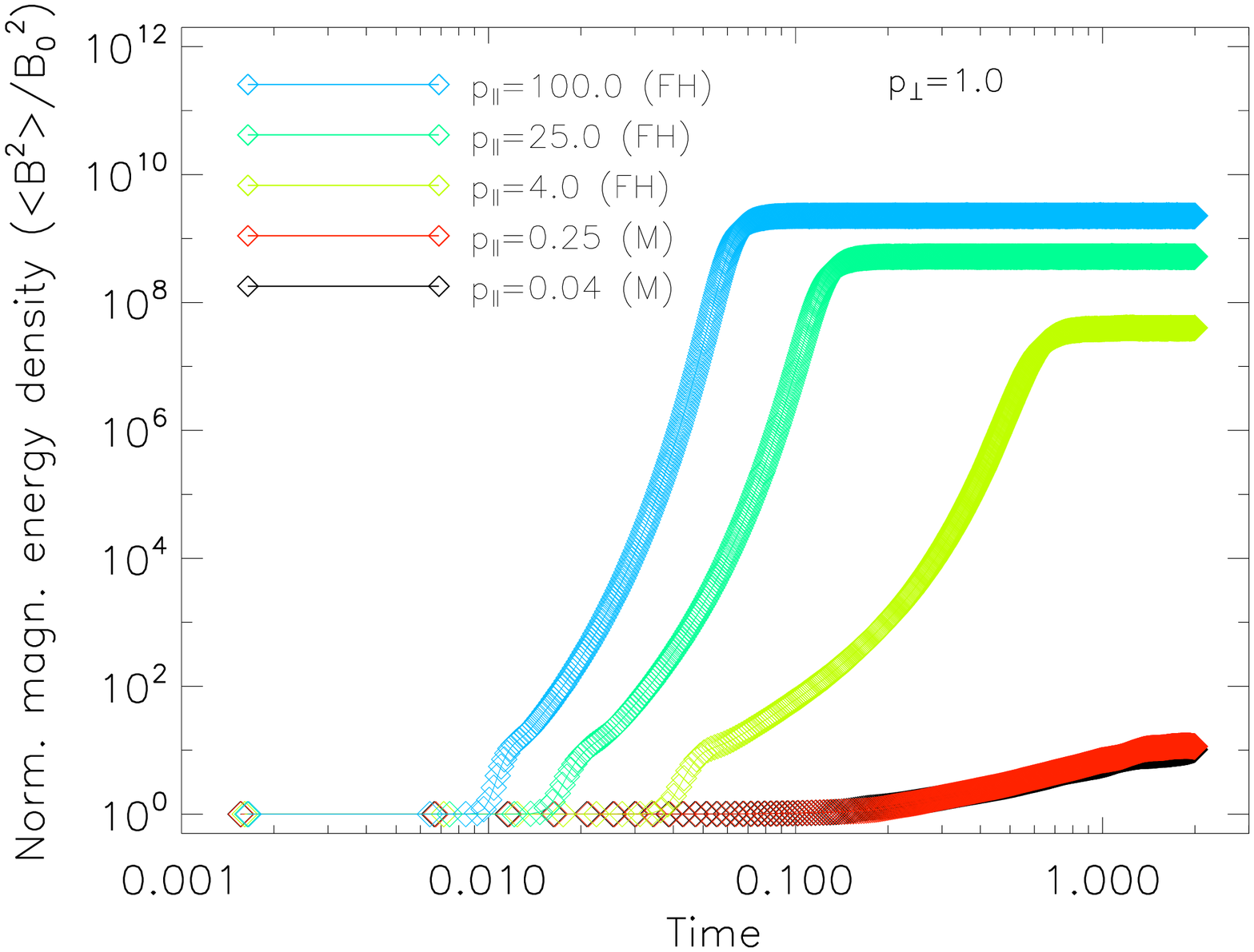}
\includegraphics[width=8cm]{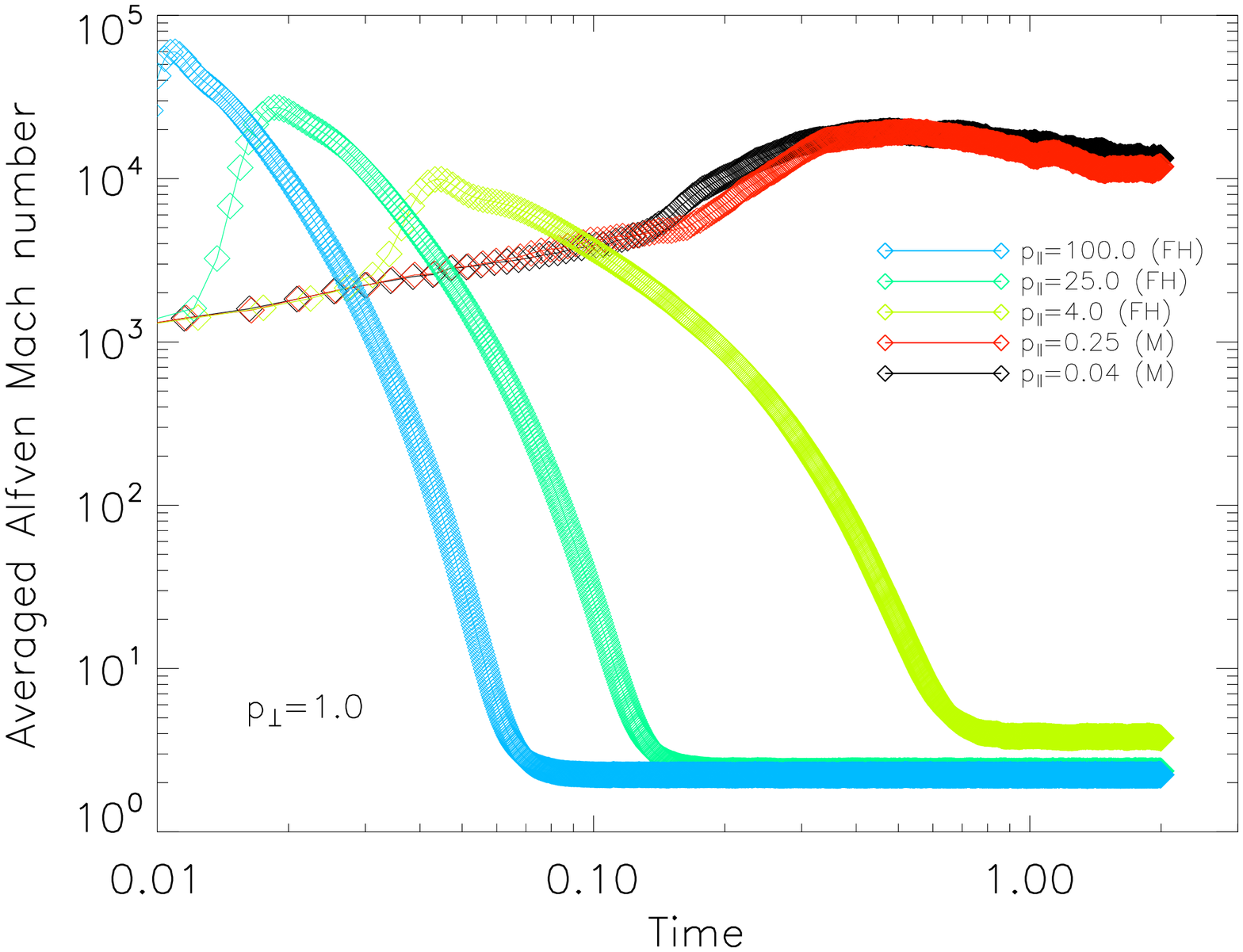}
\caption{The time evolution of the magnetic energy density normalized by its
initial value (left) and of the Alfv\'enic Mach number (right). The data shown
correspond to the models with $p_\perp = 1.0$, $256^3$ resolution, but
$p_\parallel = 100.0$ (blue), 25.0 (green), 4.0 (yellow), 0.25 (red) and 0.04
(black).}
\label{fig_anisotropy}
\end{figure*}

\begin{figure*}
\centering
\includegraphics[width=8cm]{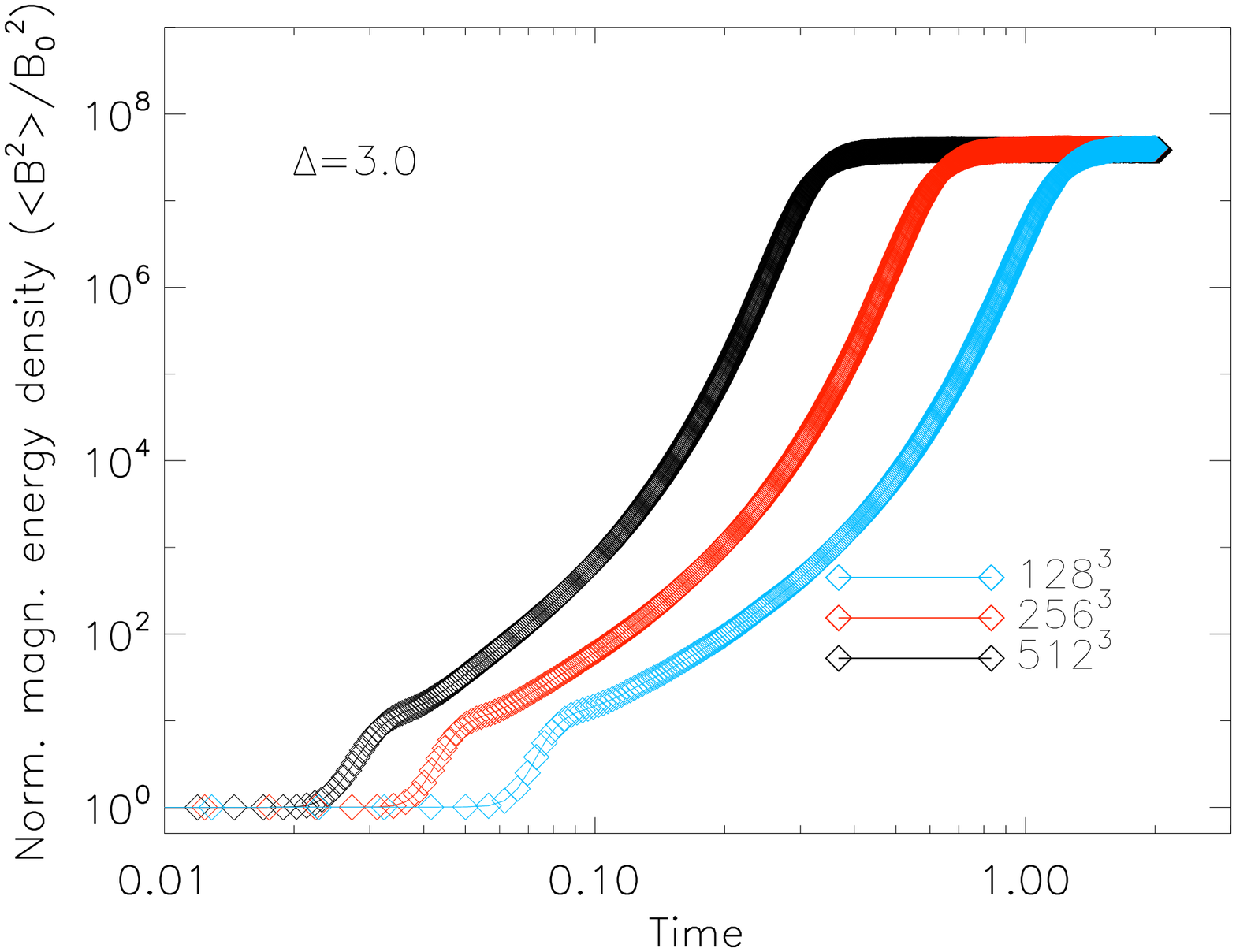}
\includegraphics[width=8cm]{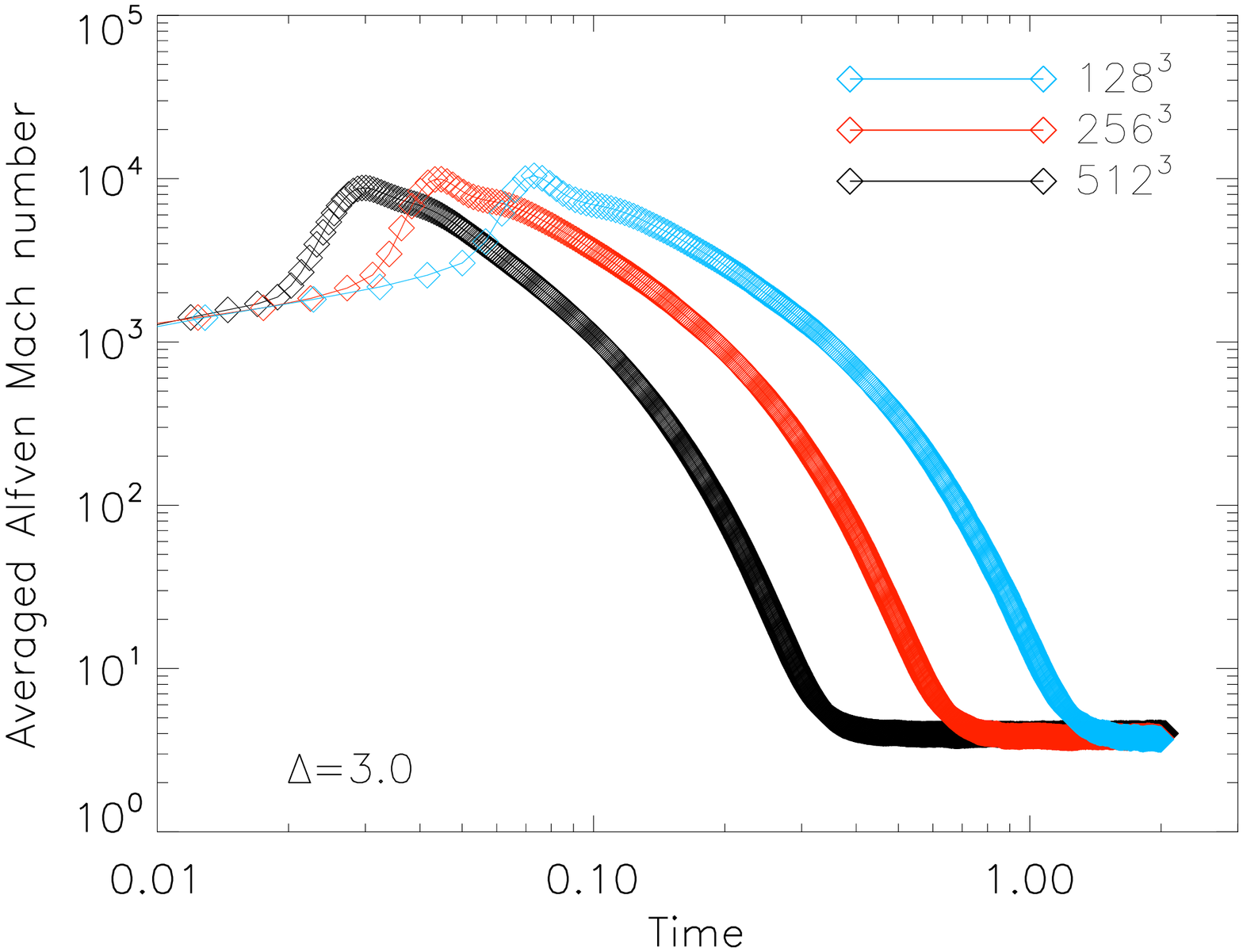}
\caption{Same as Figure \ref{fig_anisotropy} but for the models with $\Delta =
3.0$,  $\beta_0 = 2 \times 10^8$, and different resolutions: $128^3$ (blue),
$256^3$ (red) and $512^3$ (black).}
\label{fig_resolution}
\end{figure*}

\begin{figure*}
\centering
\includegraphics[width=8cm]{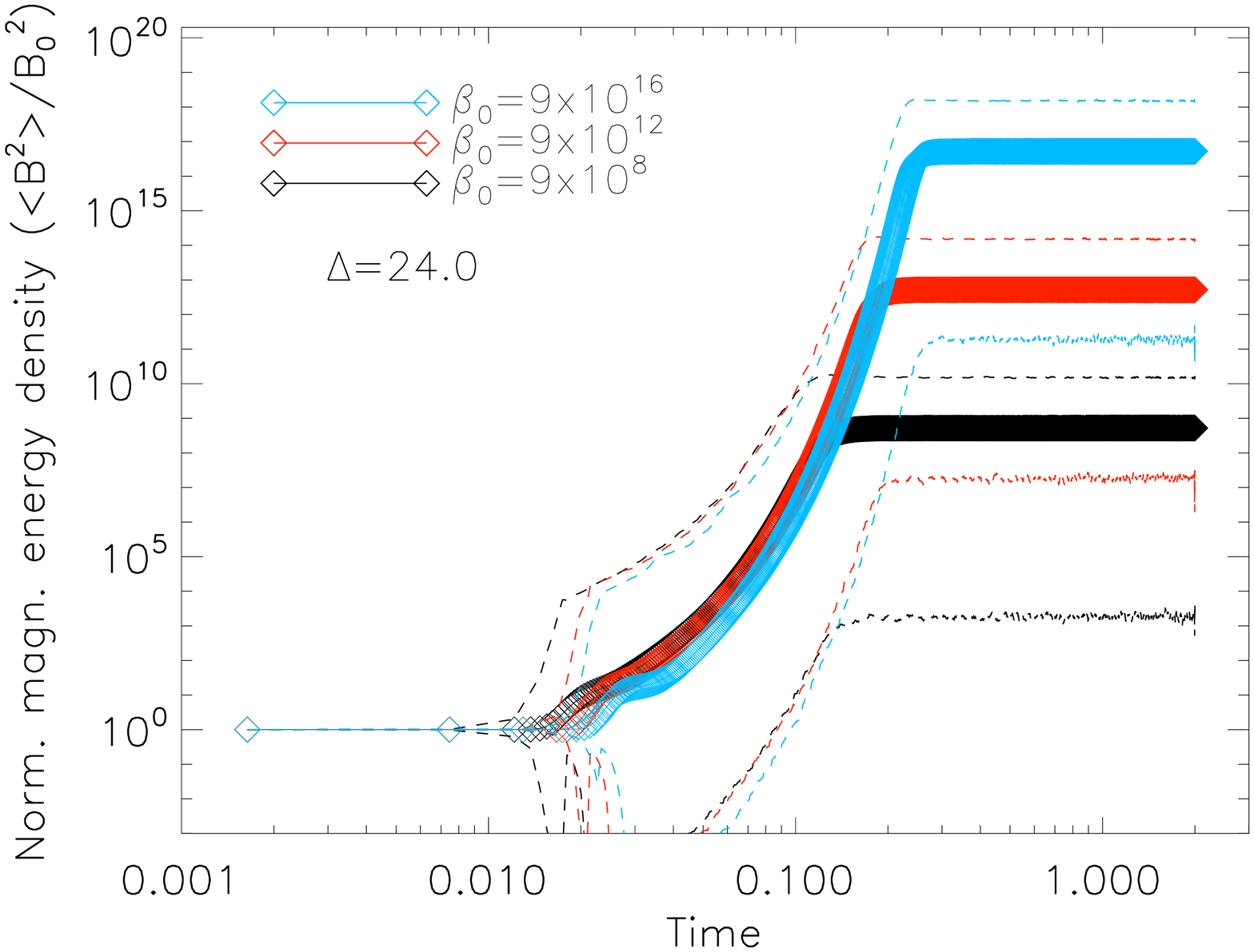}
\includegraphics[width=8cm]{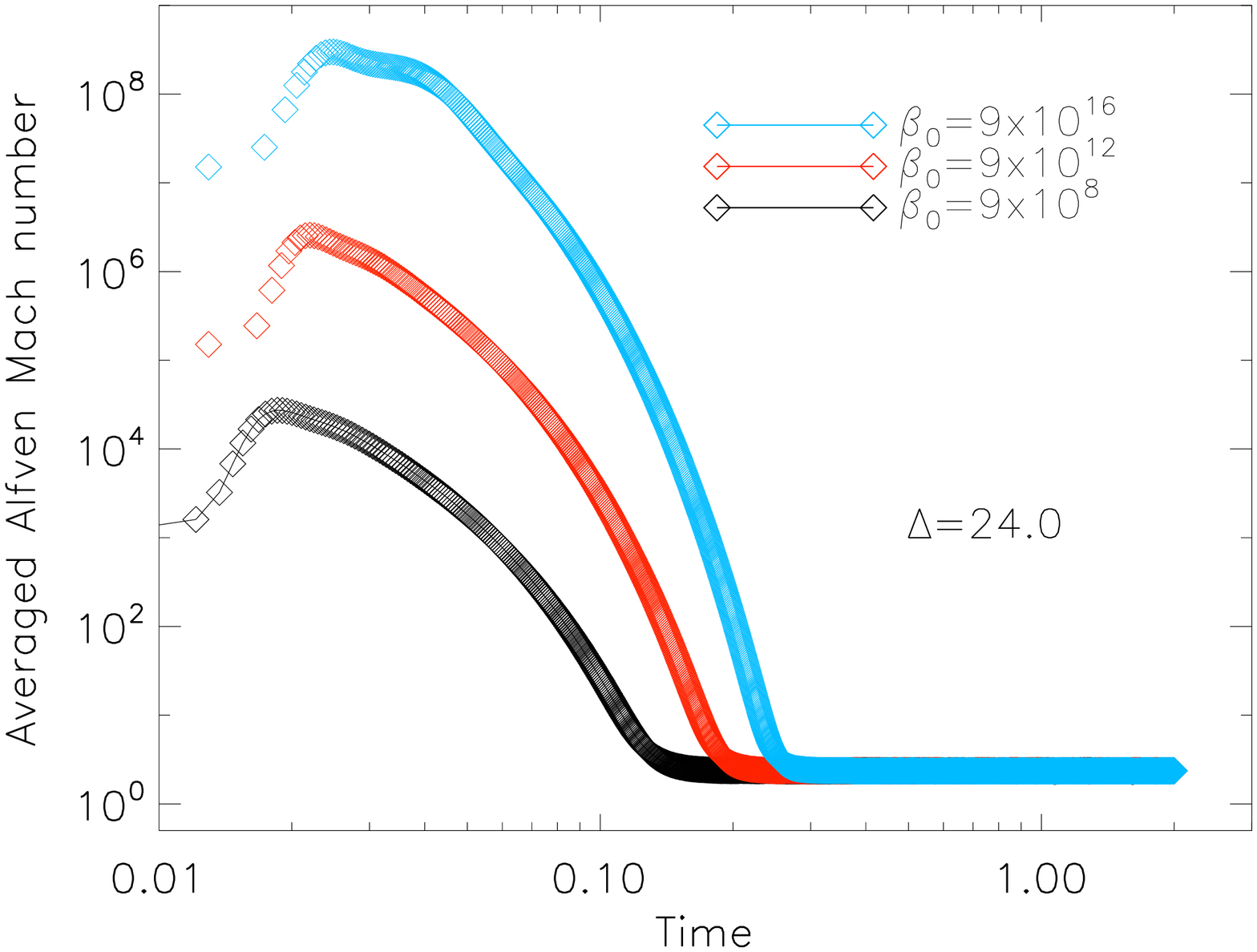}
\caption{Same as Figure \ref{fig_anisotropy} but for the models with $\Delta =
24.0$,  $256^3$ resolution, and different seed fields: $\beta_0=7\times10^{16}$
(blue), $\beta_0=9\times10^{12}$ (red) and $\beta_0=9\times10^8$ (black).}
\label{fig_seed}
\end{figure*}

As indicated by the power spectra shown in Fig.\ref{fig_spectra}, the firehose
unstable models result in amplified magnetic field fluctuations. In the left
panel of Fig.\ref{fig_anisotropy} we present time evolution of the magnetic
energy density for models with different pressure anisotropy ratios. The
magnetic energy density was averaged over all grid cells of the simulated box.
As mentioned above, the mirror unstable models show moderate growth rates for
the magnetic field perturbations, which occur at timescales of $\sim t_{\rm
dyn}$, in agreement with a typical slow turbulent dynamo. The firehose unstable
models on the other hand present much larger amplification values, corresponding
to $> 7$ orders of magnitude for the three models shown. Also, the amplification
timescales are much shorter than in the mirror unstable case, with $\tau_{\rm
gr} \ll t_{\rm dyn}$. For these models the saturation of magnetic energy density
occurs at larger levels for larger pressure anisotropies, as expected for the
saturation threshold $\beta_{\rm sat}^{-1} \propto |\Delta|$ (see
Eq.\ref{bevol}). Also, in qualitative agreement with Eq.\ref{growthrate}, the
saturation timescales for the magnetic field amplification decreases with
pressure anisotropy. The timescales and amplification levels for saturation are
also presented in Table \ref{tab:sims}.

Another interesting feature related to the time evolution of the magnetic energy
density is the presence of a ``knee'' separating two different growth regimes.
The energy density level of the knee is similar for all models. As pointed in
Section 3, if the turbulent rate of strain is larger than the growth rate of the
instabilities the turbulent dynamo regime dominates. This is precisely the
physical motivation of the knee. At the early stages turbulence is driven at
large scales, which are related to small growth rates of the instabilities. At
this point Eq.\ref{dyn1} is not yet valid given that the turbulence is not fully
evolved, but could be replaced with a reduced effective $Re$ number instead due
to the shorter inertial range. The actual magnetic growth rate, related to the
rate of strain, increases with time as the cascade develops and the inertial
range grows. This behavior is also observed in the curves of magnetic energy
density evolution. Once perturbations grow at a particular scale, at which the
growth rate of the instabilities take over the numerical dissipation, the
explosive amplification takes place.

The right panel of Fig.\ref{fig_anisotropy} presents the time evolution of the
turbulent Alfv\'enic Mach number for the different models. All models start with
similar turbulent strengths. From this it is clear that, for the cases with
$\Delta<0$, the bulk of the magnetic field amplification is not related to the
turbulence itself (though it is needed to seed the fluctuations that are subject
to the instabilities at a later stage). The mirror unstable models demonstrate
very similar evolutions while, on the other hand, the firehose unstable models
show different evolution depending on the level of pressure anisotropy. Again,
as pointed before, the larger the pressure anisotropy the faster the plasma
reaches saturation.

The same study is presented in Fig. \ref{fig_resolution}, but for the different
numerical resolutions. We can observe that the evolution of the average magnetic
energy density and the Alfv\'enic Mach number present similar profiles for the
different models. Therefore, the overall time dependency function of these
quantities is well converged in our models. The convergence is good for the
asymptotic value of the average magnetic energy density as well, which saturates
for similar values regarless the resolution and depends on $\Delta$ exclusively.
The timescales for saturation however is resolution dependent, as we have
pointed previously, being shorter for higher resolutions. This is in agreement
to the fact that finer resolutions allow the growth of the shorter wavelength
perturbations, which have faster growth rates.

The effect of the seed field intensity on the magnetic growth rate is shown in
Fig.\ref{fig_seed}, for the models with $\Delta = 24.0$  and numerical
resolution of $256^3$ cells. Here the different seed field intensities
correspond to initial thermal to magnetic pressure ratios
$\beta_0=9\times10^{12}$ (red) and $\beta_0=9\times10^8$ (black).  The time
evolution of the Alfv\'enic Mach number (right panel) shows that saturation
occurs at similar levels regardless of the seed field intensity. The saturation
occurs once instabilities drive the amplification to $\sim |\Delta|p_\perp$
level, as predicted from the CGL-MHD stability condition. Weaker initial
magnetic fields result in later saturation timescales. The delay, compared to
the models with stronger initial fields, occurs as the road up to the saturation
level must take longer for weaker seed fields.

Notice that the ``knees", i.e. the transition between the turbulent and
instability driven dynamos, occur at similar times
($t \sim 0.02 - 0.03 t_{\rm dyn}$) regardless the initial
magnetic field intensity. Also, the magnetic field amplification, relative to
the seed field intensity, of the knee is similar in the three models. Both
features are in agreement to Eq.\ref{dyn1}. That equation reveals that, for the
same timescales and similar turbulent properties, the relative amplification
$\Delta B/B$ should be similar for all models.

The time evolution of the minimum and maximum magnetic energy densities are also
shown in the left panel of Fig.\ref{fig_seed}, as dashed lines. These follow
qualitatively the same profiles of the mean magnetic energy density, except for
the short period before the ``knee'', when the maximum energy increases while
the minimum energy decreases with time. This process is due to the turbulence
that increases the dispersion of the magnetic field distribution, before the
growth of the instabilities. The second phase happens when the instability
growth rate becomes more important than the turbulent rate of strain, therefore
the ``knee'' occurs earlier for the maximum value curve (larger rate of strain),
and later for the minimum value curve (lower rate of strain).

\begin{figure*}
\centering
\includegraphics[width=8cm]{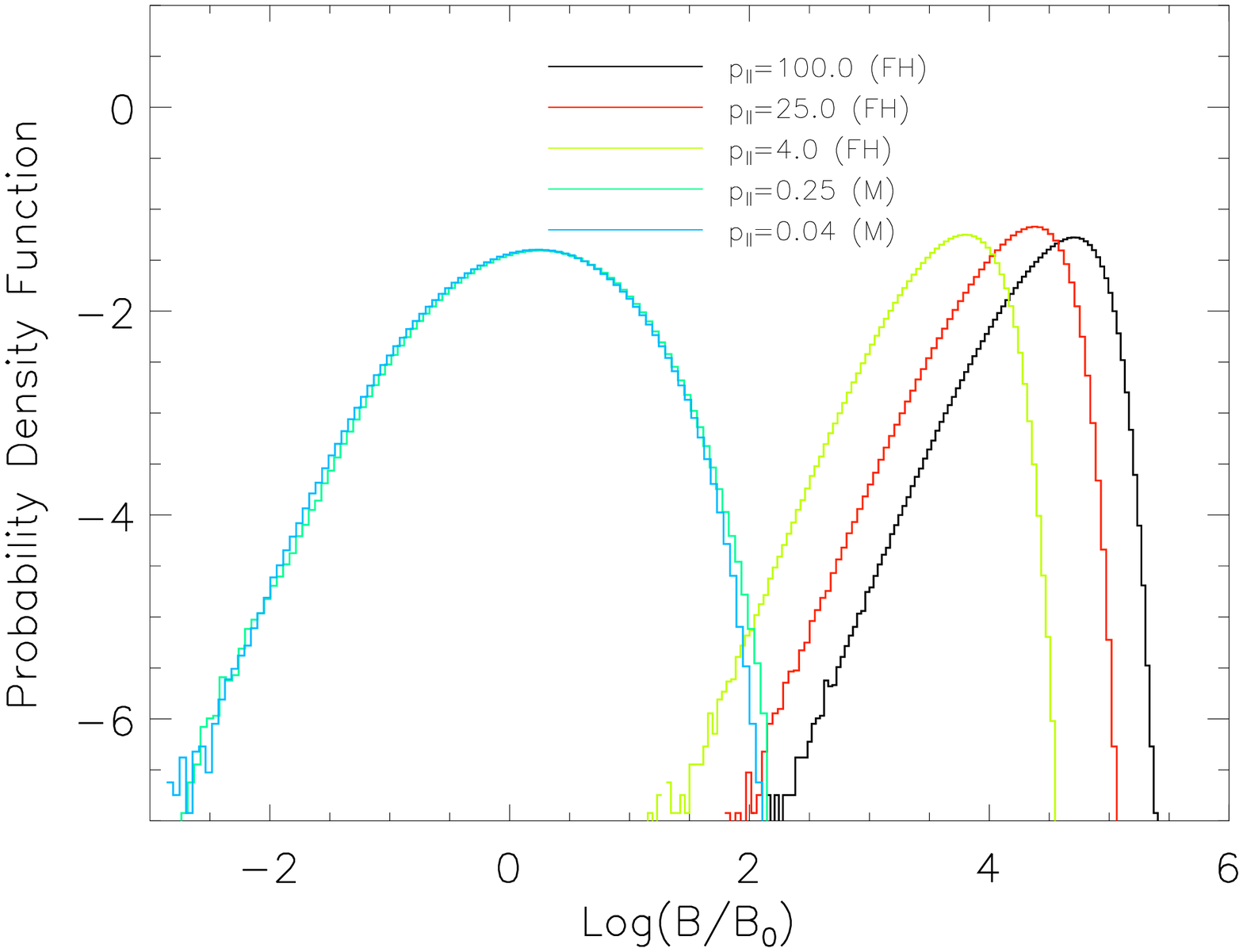}
\includegraphics[width=8cm]{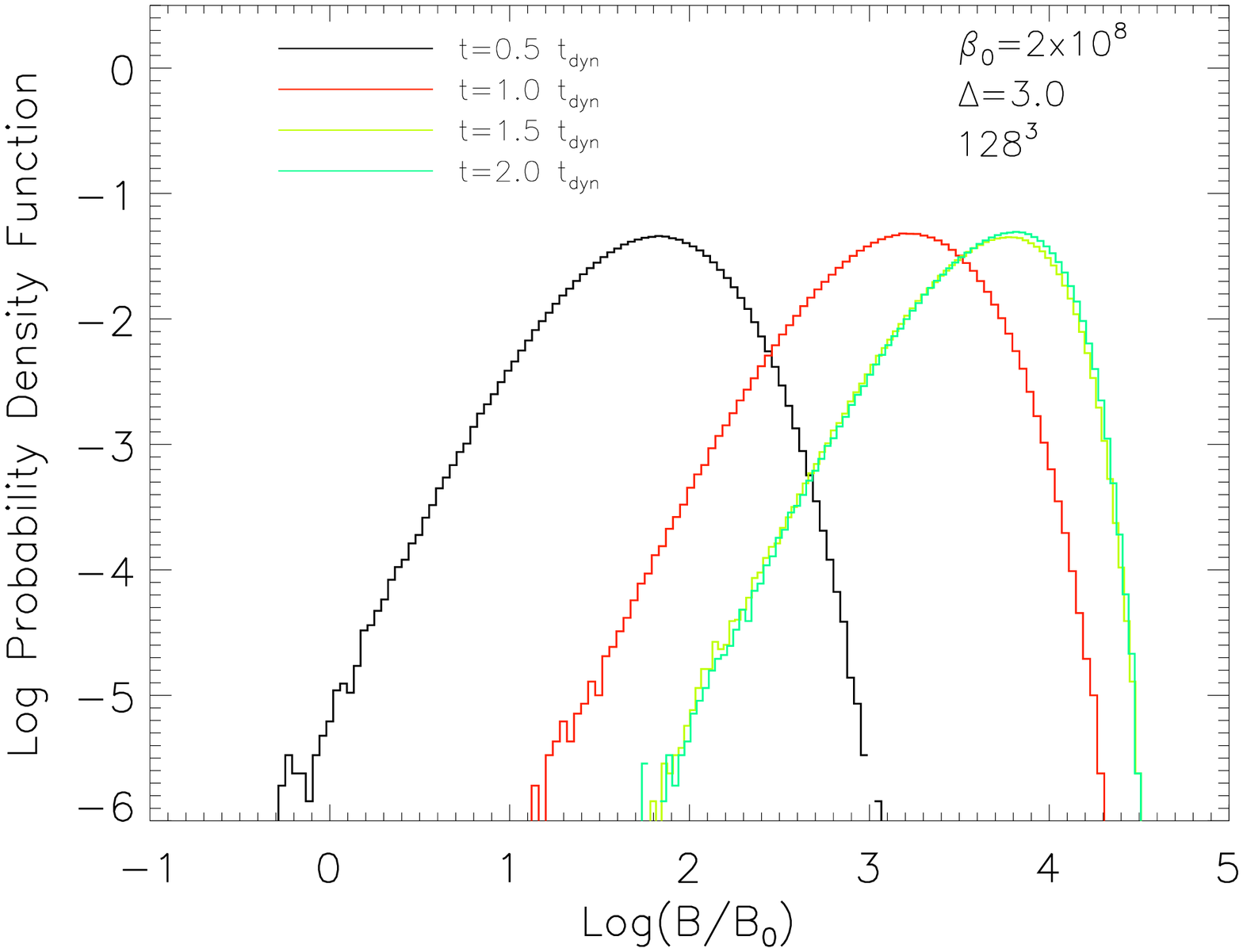}
\caption{Probability distribution functions (PDFs) of the magnetic field
intensity, normalized by its initial value, computed from all grid cells. {\it
Left:} for the models with different pressure anisotropies, at $2.0t_{\rm dyn}$.
The data represent the models with $p_\perp = 1.0$, $256^3$ resolution, but
$p_\parallel = 100.0$ (black), 25.0 (red), 4.0 (yellow), 0.25 (green) and 0.04
(blue). {\it Right:} for the $\beta_0 = 2 \times 10^8$, $\Delta=3.0$ and $128^3$
resolution model at the different times: $t=0.5$, 1.0, 1.5 and $2.0t_{\rm
dyn}$.}
\label{fig_pdfs}
\end{figure*}

The turbulent broadening of the magnetic energy density is clearly seen in the
probability distribution function (PDF) shown in Fig.\ref{fig_pdfs}. In the left
panel we present the PDFs as they depend on the pressure anisotropy. We show the
models with $p_\perp = 1.0$, $256^3$ resolution, for $p_\parallel = 100.0$
(black), 25.0 (red), 4.0 (yellow), 0.25 (green) and 0.04 (blue), all at
$2.0t_{\rm dyn}$. All models start with a delta function PDF, at $B/B_0=1.0$.
The broadening observed in the mirror unstable models exemplifies the role of
the injected turbulence on the distribution of the magnetic energy over the
simulated domain. There is clearly no net amplification of the averaged field
intensity though. The firehose unstable models present PDFs that are shifted and
skewed. Not only the averaged field intensity is amplified, as discussed above,
but these PDFs - as a whole - are shifted towards larger intensities, being the
shift proportional to the pressure anisotropy.

The skewness of the PDFs arises as their peaks are further shifted towards
larger intensities, compared to the bulk of the distribution. This occurs if the
amplification of the magnetic field is larger/faster for stronger fields. Such
behavior could be understood as a transient process due to the turbulent
broadening, in which turbulence would naturally build-up strong field in more
regions as time evolves, but this is not the case. The mirror unstable models do
not present similar skewness. Also, the time evolution of the PDFs do not
support such possibility. In the right panel of Fig.\ref{fig_pdfs} we show the
time evolution of the PDF of magnetic field intensity for a single model. We
have selected the model with $128^3$ resolution because it presents the slower
evolution of the distribution, and the PDFs obtained at different snapshots can
be compared. Here we show four PDFs calculated at $t=0.5$, 1.0, 1.5 and
$2.0t_{\rm dyn}$. The general PDF profiles change slightly with time, but is
similarly skewed from $t=0.5t_{\rm dyn}$ up to saturation time. The aparent
constant skewness of these curves shows that the changes in the distribution of
magnetic field intensity occurs too early to be accounted for the turbulence,
and should be related to the instability itself. This is in agreement with
Eq.\ref{bevol}, which predicts faster amplification for stronger fields.

\subsection{Comparison to the analytical approximations}

Obviously, any comparison between the simulations presented here and the
magnetization of the intergalactic medium needs caution. The numerical
simulations presented here correspond to full three-dimensional and stochastic
numerical solutions of the problem. It represents a good benchmark for
the analytical estimates (and approximations related) of the simplest
zero-dimension, constant-$\Delta$, $\beta \rightarrow \infty$ limit, derived in
Section 3.

\begin{figure}
\centering
\includegraphics[width=8cm]{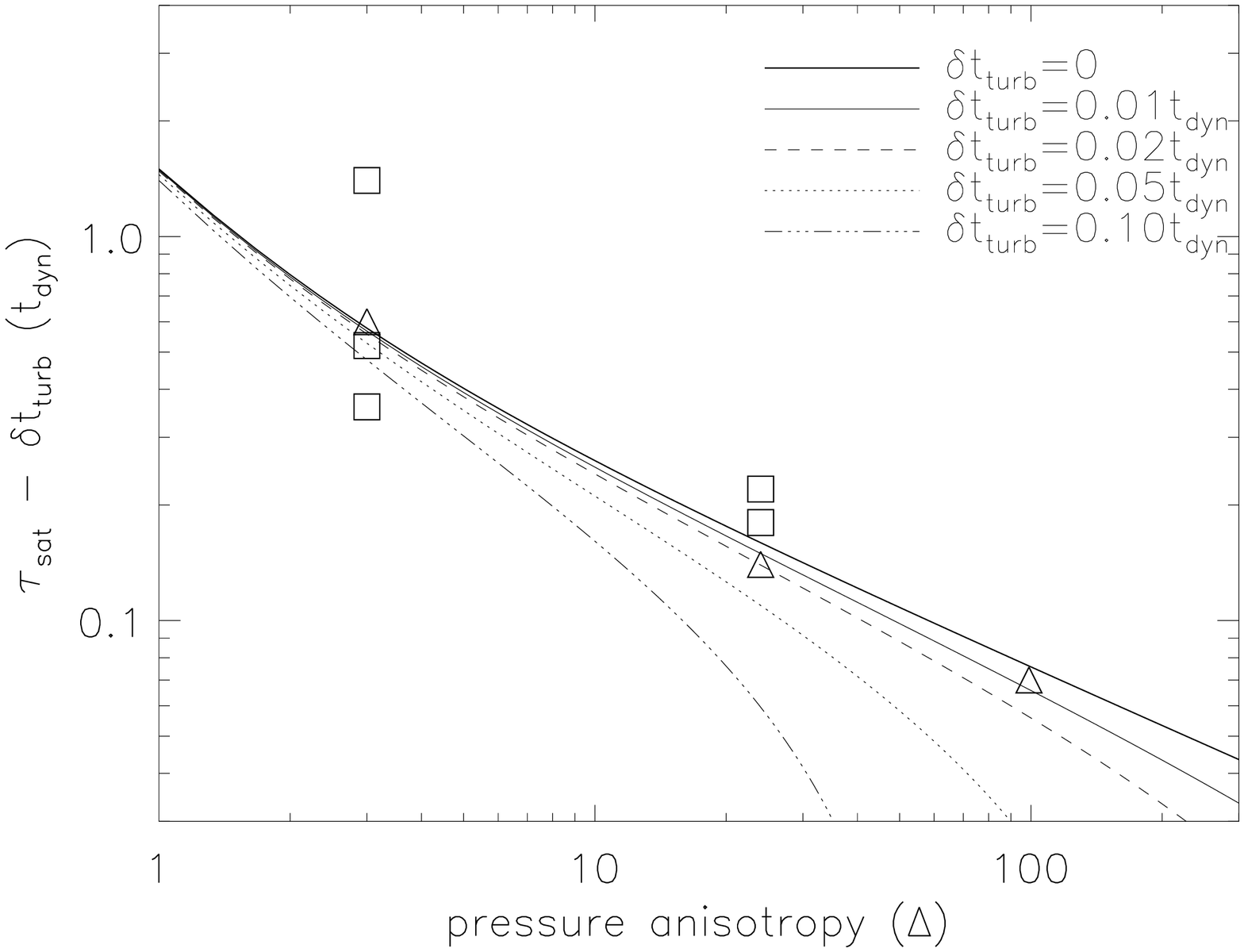}
\includegraphics[width=8cm]{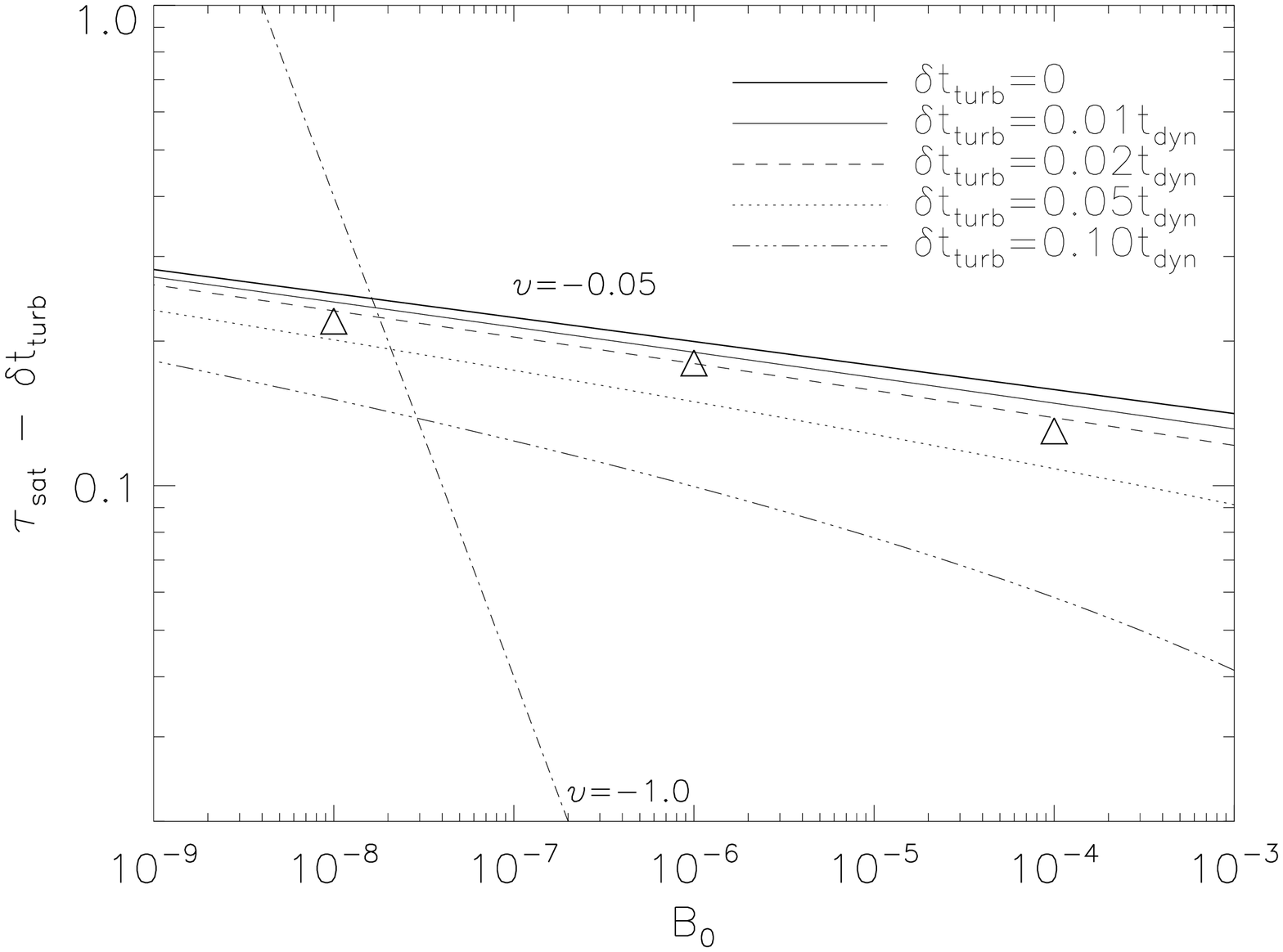}
\caption{Comparison between the numerical simulations and the analytical
predictions given in Section 3. {\it Top}: $\tau_{\rm gr} \ vs \ \Delta$
correlation. Triangles correspond to the simulations with similar initial
conditions except for the pressure anisotropy ($\Delta=3, 24$ and $99$), which
are the ones used for the correlation analysis. The lines correspond to the
solution of Eq.\ref{bevol} for different turbulent delays $\delta t_{\rm turb}$
(see text). {\it Bottom}: $\tau_{\rm gr} \ vs \ B_0$ correlation. Again,
triangle represent the models with similar initial conditions, but now with
different seed field intensities ($\beta_0=9 \times 10^8$, $9 \times 10^{12}$,
and $9 \times 10^{16}$). The lines correspond to correlations $\tau_{\rm gr}
\propto B_0^{\vartheta}$, being Eq.\ref{growthrate} the case with
$\vartheta=-1.0$.}
\label{fig:an_vs_num}
\end{figure}

We start analysing the dependency of the saturation timescale observed in the
simulations with the pressure anisotropy ratio ($\Delta$). In the top panel of
Fig.\ref{fig:an_vs_num} we present the values obtained from the simulations, as
described in Table \ref{tab:sims}, together with a number of analytical
solutions for comparison, as given by Eq.\ref{bevol}. The triangles represent
the three models with similar parameters except for the pressure anisotropy,
therefore we fit the analytical solutions to these data. The squares correspond
to the models with different initial conditions and are included for the sake of
comparison only. Here we also consider the fact that the simulated instabilities
are slightly delayed by the fact that we do not initiate the models with fully
developed turbulence. On the contrary, the initial configuration is that of an
uniform magnetic field, which is disturbed as turbulence injected at large scale
cascades. This turbulent delay ($\delta t_{\rm turb}$) is introduced as a free
constant parameter, i.e. $\tau_{\rm gr} \sim t_{\rm sat} - \delta t_{\rm turb}$.
Overall there is reasonable agreement between the simulations and the analytical
solutions, being those with $0.01t_{\rm dyn} < \delta t_{\rm turb} < 0.05t_{\rm
dyn}$ particularly good. It is interesting to note that this range is similar to
the timescales of the ``knees'' observed in Fig.\ref{fig_anisotropy}.

In the lower panel of Fig.\ref{fig:an_vs_num} the same study is presented but
now for the initial values of the seed fields. Here the triangles represent the
three models with similar initial conditions except for the seed field
($\beta_0=9 \times 10^8$, $9 \times 10^{12}$, and $9 \times 10^{16}$). According
to Eq.\ref{growthrate}, the best fit to the data should occur for $\tau_{\rm gr}
\propto B_0^{\vartheta}$, with $\vartheta=-1.0$. However, as clearly seen from
the plot, the best fits occur for a much weaker dependence (or even
no-dependence) of $\tau_{\rm gr}$ with the initial seed field, being $\vartheta
\sim -0.05$.
The apparent discrepancy between the two results is not real though. Notice that
Eq.\ref{growthrate} is derived as the magnetic field evolution once the
condition $\nu_{\rm ii} \ll \nu_{\rm scatt}$ is fulfilled. The term $B_0$ is
therefore somewhat misleading in the sense that one must consider the magnetic
field intensity once $\nu_{\rm ii} \ll \nu_{\rm scatt}$, i.e. a transition field
intensity. For the analytical model the transition field intensity is determined
by $\nu_{\rm ii}$, while in the simulations by the numerical dissipation, i.e.
in both cases it is insensitive to initial magnetic field. Another major effect 
resposible for this discrepancy is that fact that the growthrate depends on 
the Larmor radius, which cannot be propely modelled in the simulations. Naturally, 
$\tau_{\rm gr}$ obtained from the simulations is overestimated as $B$ increases with time. 

\section{Discussion}

Several authors have already discussed the effects of turbulence \citep[see e.g.][and many others]{banerjee04,saveliev12, iapichino12, saveliev13,cho14} and structure formation \citep[see e.g.][and many others]{dolag02,vazza14} on the evolution of the intergalactic magnetic field. The lack of detection of TeV 
radiation from blazars points towards the existence of amplified magnetic fields, with filling factors 
around 60\%, even at the intergalactic medium voids \citep{nero10,dolag11}. For this reason, turbulence - as a ubiquitous phenomenon - has 
been preferred as main mechanism for field amplification. The 
common problem related to turbulent models of collisonal plasmas is the typical timescale 
needed for magnetic field amplification. 
As explained previously the timescales associated to turbulent amplifications are 
too large for the magnetic seed amplitudes suggested so far. Structure formation, on the other hand, would be dominant (due to shear, compression or other star-formation related mechanisms) in further amplification of the fields.

In this work we explore a possible scenario for magnetic field amplification
at high redshits (post-recombination), based on the growth of collisionless plasma instabilities.
Two similar works have dealt with the problem in a similar approach, being one
analytical \citep{sche06a} \citep[see also][]{sche06b}, and the other purely
numerical \citep{lima14}. In \citep{sche06a} an analytical simplification of the
problem is posed, which inspired most of the analytical derivation presented
here as well. In their work, however, it is assumed that both, the damping at
small scales and the pressure anisotropy $\Delta$, are similarly related to
$\nu_{\rm scat}$ and $\nu_{\rm ii}$. This in order to obtain the time evolution
of $\Delta$ and $B$ in terms of an effective collision rate. In the present we
do not constrain $\nu_{\rm scat}$ which is, in another hand, implicitly
constrained to the assumption that $\Delta$ is kept constant during the
evolution of the system. Interestingly, for the super-exponential phase, both
works predict the magnetic field intensity growth rate as $\propto
(1-t/t_{c})^{-b}$, though with different slopes and characteristic timescales.
In both estimates the growth of the magnetic field by collisionless
instabilities could, in principle, explain the magnetization of the local
Universe based on very weak seeds ($< 10^{-17}$G). Only the analytical model
presented in this work has been compared to numerical simulations.

With respect to the numerical solutions, in \citet{kowal11} the authors
performed simulations with strongly magnetized plasmas  and could not address
the amplification problem. In this sense, the numerical study presented here
stands as an extension of that particular work but focused on the statistical
properties of the magnetic field and the role of collisionless plasma
instabilities on the amplification of the magnetic field. The scalings found for
density and velocity fields in the firehose unstable models of that work
resemble those obtained here for the magnetic field, with a power spectrum peak
at the smallest scales available in the system.
Similar statistical properties were obtained by \citet{lima14}, who performed
collisionless plasma simulation with variable pressure anisotropy and studied
the statistical properties of the magnetic fields as well. The plasma properties
were studied in a variable-$\Delta$ framework though, with isotropization
mechanisms being introduced {\it ad hoc}. Still, it was shown that even in the
cases where isotropization is strong enough to wash out the imprints of
instabilities in the statistical properties of the plasma, their dynamical
effects are important locally. The amplification of the magnetic field was
studied and, as a consequence of small volume coverage of unstable regions,
their general conclusion was that the instabilities were not efficient.
Differently to the assumption made in this work, their models do not include an
external source term to excite the pressure anisotropy, which could - in
principle - change the system to a quasi constant-$\Delta$ case, similar to
ours.

Possible observational evidence (or constraints) for the process described in this work could 
be obtained from the cosmic microwave background. If collisionality condition was satisfied at pos-equipartition 
era, the magnetic field at small scales could have been amplified prior to the recombination era, and 
the ionization structure of the Universe would be different, with effects on Silk damping \citep[see][]{jeda13}. On the other hand, if the plasma becomes gyrotropic at $z<z_{\rm rec}$, such effect would not be relevant.

The main problem with the models mentioned above is the lack of a proper
physical description of the isotropization of the particles momenta as the
system evolves. Differently to the previous works we consider the balance
between pressure anisotropy sources and isotropization effects, which simplifies
dramatically the solution of the problem and results in a problem that can be
addressed both numerically and analytically.
Nature is obviously far more complex than a constant-$\Delta$ model, but still
any approximated model that can be studied by means of both analytical and
numerical methods may be extremely usefulon the understanding of the basic
dynamical properties related to the dynamics of collisionless plasma fluids.

\subsection{Magnetic field characteristic lengthscales}

Even if the amplification of the magnetic field, from seed fields up to
equipartition levels, could be explained by the collisionless plasma
instabilities, one must still consider its characteristic lenghtscale. The
zero-dimensional anaytical models cannot account for this properly. Still, it is
reasonable to consider the typical lengthscale to be that of maximum growth
rate. Indeed, from the numerical simulations presented in this work, it has
been shown that the turbulent flows cannot modify the magnetic field structure
in timescales at which the amplification occurs. The spectral distribution of
the magnetic field obtained from the simulations peak at the smallest scales
possible, those at which the numerical dissipation overtakes the field growth.
The finer the resolution the smaller is the characteristic lengthscale. 

In the
intergalactic medium this would be given by the natural process that can
overtake the pressure anisotropy instabilities, particle collisions. With a
typical mean free path of $\sim$ few kpc, one should expect the magnetic field
energy to be concentrated in fluctuations with approximately this lengthscale.
However, magnetic field correlation lengths as large as $\sim$Mpc has been inferred from
observations \citep[e.g.][]{ferreti2012}, posing the challenge not only on how to amplify
the magnetic field but also on how to distribute the energy from the small to the
large scales. In our firehose unstable CGL-MHD models, we obtained power spectra which are steeper than the expected for the kinetic phase of the small-scale dynamo. This demonstrates that the instabilities are indeed the dominant process in place in these models. On the other hand, depending on the level of pressure anisotropy it also results in weaker large scale fields, if compared to the SSD estimates.  
Given that most of the energy of the magnetic 
fluctuations are placed at the smallest scale in the gyrotropic fluid approximation (i.e. $\lambda_{\rm mfp}$), 
one could estimate (see Section 3) the magnetic field intensity at large scales as 
$B_L \sim \left(|\Delta|p_\perp \right)^{1/2} \left(\lambda_{\rm mfp}L^{-1}\right)$. For $n\simeq 10^{-3}$cm$^{-3}$, 
$kT\simeq1$eV, one finds $B_{\rm Mpc} \sim \sqrt{\Delta} 10^{-10}$G.

The answer to this problem may be standing in the properties of the turbulence itself. As
discussed earlier in this paper, turbulence stretches and folds the magnetic
field in a process that drives the turbulent dynamo. This process is dominant at
small scales, and is quenched in a given scale when the magnetic energy reaches
equipartition with the turbulent one. In the collisionless plasma the
instabilities are responsible for the growth of magnetic field at the small
scales. Even though turbulence cannot stretch the field lines at the smallest
scales it could, in principle, do it at the largest ones depending on its
strength. This is clearly shown in Figs.\ref{fig_anisotropy},
\ref{fig_resolution} and \ref{fig_seed}, where the evolution of the turbulent
Alfv\'enic Mach number is always larger than unity, even after the saturation of
the magnetic field. This because the velocity dispersion is dominanted by the
large scale eddies, where equipartition was not reached. For these particular
models, the turbulence is responsible for transfering the magnetic field energy
from small to large scales. This process should be even more important if the
pressure anisotropy is quenched (this is not the case of the simulations, in
which we have kept it constant). The quenching of the pressure anisotropy,
followed by the transfer of magnetic energy from small to large scales by
super-Alfv\'enic turbulence could therefore explain the observations. In 
other words, as shown in the numerical simulations, the CGL-MHD instabilities are presumably 
dominant over the kinematic phase of the small-scale dynamo, but possibly not 
for the later nonlinear phase (once the instabilities are quenched).

Another possible solution for this problem would be related to the formation of 
 large structures of matter (e.g. massive galaxies, groups and clusters). It is 
 quite clear now that the large scale magnetic field is not primordial, but has constantly 
 evolved with the dynamical evolution of Universe \citep[][]{jeda11}.
This was also pointed by \citet[][cf. references therein]{schlickeiser05}, in the context of small-scale fluctuations of Weibel instability. The collapse of gas into galaxies or clusters \citep{vazza14}, as well as large-scale shears, could in principle amplify coherent components of the magnetic field. If correct, in such scenario 
magnetic energy would have been provided dominantly at small scales, in the post-recombination era, but 
reasembled into large scale coherent structures ($>100$kpc) following the recent structure formation of the dark and baryonic matter ($z<10$).

\section{Summary}

In this work we studied the magnetic field amplification process in turbulent
collisionless plasmas. In such plasmas, parallel and perpendicular pressures
(with respect to the local magnetic field) may become so anisotropic that
instabilities may become dynamically important. It has been pointed in previous
analytical models that such instabilities, acting together with a turbulent
background, may accelerate the amplification of the magnetic field up to near
equipartition levels. Such models would then have dramatic implications on our
understanding of the magnetization of the early Universe (post-recombination), 
given that the intergalactic medium then becomes collisionless. These models provide timescales
and saturation estimates for the magnetic energy density, but have never been
confronted by full three-dimensional numerical simulations. The aims of this
work were to revisit the analytical zero-dimensional models of the magnetic
field evolution in turbulent collisionless plasmas and to provide a number of
full three-dimensional numerical simulations that could be directly compared to
these analytical estimates.

As long as pressure anisotropy is kept constant during the evolution of the
system a novel analytical solution is provided for the magnetic field evolution
in systems subject to firehose instability. As discussed before, such assumption
is not too far from reality, given that different processes act together to
increase and decrease pressure anisotropy, and some sort of equilibrium value
for $\Delta$, in unstable regime, may exist (specially for weakly magnetized
plasmas). We find that:

\begin{itemize}

\item for very weak initial seed fields, the early evolution of the magnetic
field is dominated by collisions, and the turbulent dynamo should be responsible
for the initial amplification of the field;

\item once the field grows and $\Omega_i > \nu_{\rm ii}$, a transition from
an isotropic to a gyrotropic plasma occurs and
instabilities would then be driven. A fraction of the ``free-energy'' available
($|\Delta|p$) would then be transfered to the magnetic field as the system
evolves, up to saturation;

\item saturation occurs at $\beta \sim 2/|\Delta|$ value, which is reached in a
timescale $\tau_{\rm gr} \propto B_0^{-1} |\Delta|^\psi$, with $\psi = 1/2$ or
$3/2$, for $|\Delta| \gg 1$ and $\ll 1$, respectively.

\end{itemize}

This analytical model was then compared to a number of CGL-MHD numerical
simulations, with different initial and turbulent conditions, from which the
evolution of the magnetic field from the seed field, up to saturation, was
followed. The main results obtained are as follows:

\begin{itemize}

\item the explosive growth of magnetic field energy density is observed in the
firehose unstable models, being the amplification timescales approximately
proportional to $|\Delta|^\psi$, as was predicted in the simple analytical
model;

\item there is no clear dependency between $\tau_{\rm gr}$ and $B_0$ in the
simulations, in contradiction to the linear anti-correlation derived in the
analytical model. This is explained in terms of $B_0$ being the field intensity
at $\Omega \sim \nu_{\rm ii}$ transition, rather than the initial seed value. Also, 
because the code is unable of evolving the Larmor frequency as a function of $B$;

\item the power spectrum of the magnetic field peaks at the smalles scales not
dominated by the numerical diffusion, and the transfer of energy to large scales
must occur aftewards, by turbulence itself, once saturation occurs.

\end{itemize}

\acknowledgments
The authors would like to acknowledge the anonimous referee for the useful comments and suggestions that 
helped improving the current manuscript. We are also very thankful to Alex Schekochihin for valuable discussions 
about the content of this paper. DFG thanks the European Research Council (ADG-2011 ECOGAL), and
Brazilian agencies CNPq (no. 302949/2014-3), CAPES (3400-13-1)
and FAPESP (no.2013/10559-5) for financial support. GK acknowledges
support from FAPESP (grants no. 2013/04073-2 and 2013/18815-0).

\end{document}